\documentclass[showpacs,aps,prd,preprint,nofootinbib,showkeys,unsortedaddress,raggedbottom]{revtex4-1}
\pdfoutput=1

\usepackage{bm}
\usepackage{amsmath}
\usepackage{graphicx}
\graphicspath{{figures/}}
\usepackage{subfigure}
\usepackage{float}
\usepackage[usenames,dvipsnames]{color}
\definecolor{darkblue}{RGB}{0,0,196}
\usepackage[colorlinks=true,linkcolor=darkblue,citecolor=darkblue,urlcolor=darkblue]{hyperref}

\usepackage{setspace}

\usepackage{footmisc}

\usepackage[makeroom]{cancel}

\usepackage[utf8]{inputenc}

\DeclareMathOperator{\sech}{sech}

\def\be{\begin{equation}}
\def\ee{\end{equation}}
\def\ba{\begin{eqnarray}}
\def\ea{\end{eqnarray}}

\usepackage{lineno}

\begin{document}

\title{Dilepton production and elliptic flow from an anisotropic quark-gluon plasma}

\author{Babak S. Kasmaei and Michael Strickland}

\affiliation{Department of Physics, Kent State University, Kent, OH 44242 United States}

\begin{abstract}

\noindent
We calculate the yield and elliptic flow of mid-rapidity dileptons emitted from the quark-gluon plasma generated in Pb-Pb collisions at LHC. We use relativistic anisotropic hydrodynamics for the 3+1 dimensional evolution of the quark-gluon plasma and convolve this with the momentum-anisotropic local rest frame production rate for dileptons. The effects of momentum anisotropy of the quark distribution functions, viscosity to entropy density ratio, centrality of the collisions, and initial momentum anisotropy on the results are investigated and discussed.  

\end{abstract}

\maketitle

\section{Introduction}

Extensive phenomenological investigations of the results emerging from high energy heavy-ion collision experiments at RHIC and LHC have provided a broad but still incomplete picture of the physics of hot and dense strongly interacting matter \cite{Braun-Munzinger:2015hba, Shuryak:2014zxa, Schukraft:2017nbn, Ploskon:2018yiy}.  In particular, the success of relativistic hydrodynamical models in describing the production and azimuthal asymmetries of the produced hadrons has led to insights about the collective flow, thermodynamics, and transport properties of the expanding quark-gluon plasma (QGP) \cite{Ollitrault:2008zz, Gale:2013da, Jeon:2015dfa, Alqahtani:2017mhy}.  

The current standard approach to describe the evolution of the strongly interacting system is to consider several stages. During the initial stage ($\tau \lesssim 1 {\rm \ fm/c}$), the QGP possesses a high energy density but is driven out of equilibrium by rapid longitudinal expansion. The resulting pseudothermalized state can be used as an initial condition for the subsequent dissipative hydrodynamic evolution ($1 {\rm \ fm/c} \lesssim \tau \lesssim 10 {\rm \ fm/c}$) which dilutes and cools down the system until the final stages of hadronization, decoupling, and decay of the outcoming particles \cite{Yagi-Hatsuda}. In order to obtain information about the properties of the medium and its evolution dynamics, various observables for the nuclear collision experiments have been suggested and are used to improve both the qualitative and quantitative aspects of the models \cite{Shuryak:1978ij, Bass:1998vz, Braun-Munzinger:2015hba}.       

Electromagnetic probes have been considered as the best observables for learning information about the initial stages of the evolution of the quark matter produced in heavy-ion collision experiments, since they are not directly affected by strong interactions. 
In contrast to hadrons, electromagnetic probes such as real photons, leptons, and dileptons (virtual photons) can be generated during all stages of the evolution of the strongly interacting matter and, when produced, they can escape with much larger mean free paths than hadrons. Therefore, they may provide information about the effective temperature and momentum distribution of each stage. In particular, since the hadronic observables are mainly produced at later and colder stages, the electromagnetic probes can be considered to have the unique role of being the messengers of the initial stages of the system. However, extracting clean experimental data, constructing accurate quantitative models, and providing clear interpretation of electromagnetic probes are, in general, not straightforward tasks \cite{McLerran:1984ay, Gale:1987ki, STRICKLAND1994245, Gale:2012xq, Linnyk:2012pu, Sakaguchi:2014ewa, Bratkovskaya:2014mva, Bratkovskaya:2014toa, Endres:2016tkg, Shen:2016odt, Paquet:2017wji, Campbell:2017kbo}.  

Dileptons, in contrast to real photons, have both the invariant mass and transverse momentum as independent variables which is beneficial for comparison of theoretical models with experimental data.  Dileptons also play an important role in the study of in-medium modification of the spectral function of $\rho$ meson and its relation to the chiral symmetry restoration \cite{Rapp:1995zy, Bratkovskaya:1998pr, Rapp:1999ej, vanHees:2007th, Rapp:2016xzw}. Also, various other higher-order effects such as quantum corrections or magnetic modifications have been suggested to affect the production rate of the dileptons \cite{Weldon:1990iw, Tuchin:2013bda, Ghiglieri:2014kma, Basar:2014swa, Burnier:2015rka, Bandyopadhyay:2015wua, Hidaka:2015ima, Sadooghi:2016jyf}.

Dileptons can be produced during every stage of the evolution of the system. However, by looking at different mass or transverse momentum windows, information about different stages of the system can be separated to some extent. Lower mass dileptons ($M \lesssim 1 {\rm \ GeV}$) are believed to be mainly produced from the hadronic matter generated below the critical temperature $\sim 155 {\rm \ MeV}$. Production of intermediate mass in-medium dileptons ($1{\rm \ GeV} \lesssim M \lesssim 3{\rm \ GeV}$) is mostly affected by the partonic QGP phase before the transition to the hadronic phase.\footnote{In the intermediate mass region, besides the in-medium QGP dileptons, there are also important contributions from open heavy flavor decays. There are also contributions to hard dileptons from passage of jets \cite{Srivastava:2002ic, Turbide:2006mc, Fu:2014daa, Mukherjee:2016sep}.} Therefore, the low mass window may provide information on the spectral functions of vector mesons, while intermediate mass dileptons can give insights into properties of the earlier high temperature stage.

Most of the theoretical models of dilepton emission have provided predictions for the dilepton yield as a function of $M$ and $p_T$. In recent years, the azimuthal flow observables, e.g. elliptic flow, have provided additional information about the dilepton emission \cite{Chatterjee:2007xk, Vujanovic:2013jpa, Vujanovic:2016anq}. In particular, one can hope that, by tuning the models to both the differential yield and flow data, information about both the effective temperature and momentum anisotropy of  the early stages of the QGP can be extracted \cite{Bhattacharya:2015ada}. 

The current standard picture of collective dynamics in heavy-ion collisions suggests that the spatial asymmetry of the collisions and fluctuations in the initial shape of the system generated in heavy-ion collisions is the primary source of anisotropic collective flow observed at later stages \cite{Heinz:2013th}. In addition to the spatial anisotropy, the rapid longitudinal expansion of the system generated in ultrarelativistic heavy-ion collisions cause the local rest frame (LRF) parton-level momentum distribution functions to become anisotropic \cite{Mrowczynski:2000ed, Romatschke:2004jh}. If the deviations from LRF isotropy of the parton momentum distributions are so large that they cannot be considered as linear perturbations of isotropic distributions, one can expect to observe their effect in the emission of intermediate mass/momentum electromagnetic probes \cite{Mauricio:2007vz, Martinez:2008di, Martinez:2008mc, Ryblewski:2015hea, Bhattacharya:2015ada}. In order to check this, it is essential to have phenomenological models of electromagnetic emission from the QGP that incorporate momentum-anisotropic distributions both in the calculation of the LRF emission rates and in the space-time evolution of the fireball. 

Hydrodynamical models have been shown to be able to reproduce experimental data for the collective flow of the hadrons formed in the central rapidity region \cite{Blaizot:1990zd, Schenke:2010rr, Shen:2011eg, Alqahtani:2017jwl}. They have also been used to integrate over the space-time of the electromagnetic emission from the strongly interacting medium \cite{Paquet:2017wji}. In order to include non-ideal and non-equilibrium aspects of the QGP evolution, various generalizations of the conventional ideal hydrodynamics have been developed in recent years. Finite but small viscosity of the QGP has been incorporated in viscous hydrodynamics (vHydro) \cite{Romatschke:2007mq, Dusling:2007gi, Luzum:2008cw, Dusling:2008xj, Romatschke:2009im, Schenke:2010rr, Vujanovic:2013jpa}. Anisotropic hydrodynamics (aHydro) has been developed in order to generalize to momentum-anisotropic distributions of the QGP in local rest frame in a way that positivity of the parton distribution functions is guaranteed \cite{Strickland:2014pga}. Anisotropic dissipative hydrodynamics combines both viscous effects and momentum anisotropy \cite{Florkowski:2010cf, Ryblewski:2010bs, Martinez:2012tu, Nopoush:2014pfa, Florkowski:2016kjj}.
Previously, anisotropic hydrodynamics has been used to calculate the space-time integrated yields of QGP  dileptons \cite{Mauricio:2007vz, Martinez:2008di, Martinez:2008mc, Ryblewski:2015hea}. In previous studies along this line, the thermal dilepton yield has been calculated within the 1+1d models \cite{Mauricio:2007vz, Martinez:2008di, Martinez:2008mc}, and within 3+1d dissipative aHydro considering spheroidal momentum-anisotropic distributions in local rest frame \cite{Ryblewski:2015hea}. The elliptic flow of thermal dileptons has also been studied previously using ideal \cite{Chatterjee:2007xk, Gale:2014dfa} and viscous hydrodynamics \cite{Vujanovic:2013jpa, Vujanovic:2016anq} for the space-time evolution of the fireball.

 In this paper, we study the yield and elliptic flow of in-medium dileptons emitted from a QGP with ellipsoidally anisotropic momentum distributions. We utilize 3+1d aHydro with a realistic equation of state \cite{Alqahtani:2017mhy} to describe the evolution of the expanding QGP and convolve the space-time evolution with the anisotropic dilepton rates from the local rest frame of the fluid elements. The parameters for the background hydrodynamic evolution are tuned to reproduce soft hadron observables \citep{Alqahtani:2017jwl, Alqahtani:2017mhy}. We calculate the yield and elliptic flow only for dileptons produced in the QGP phase above the critical effective temperature. Dileptons from hadronic sources are not included in this study. Therefore, we focus mainly on intermediate dileptons with mass and transverse momentum above $1 \ {\rm GeV}$.  

\subsection{Dilepton production rate in local rest frame}

From kinetic theory, the differential production rate of thermal dileptons from the QGP is given by
\be
\frac{dN}{d^4x d^4P} = \int \frac{d^3 {\bf k}_1}{(2\pi)^3} \frac{d^3 {\bf k}_2}{(2\pi)^3}f_q({\bf k}_1)f_{\bar{q}}({\bf k}_2)v_{q\bar{q}}\sigma_{q\bar{q}}^{l^{-}l^{+}} \delta^4(k_1 + k_2 - P), \label{rate00}
\ee
where $P=(E,{\bf p})$, $k_1=(E_1,{\bf k_1})$, and $k_2=(E_2,{\bf k_2})$ are the four-momenta of the lepton pair, quarks, and anti-quarks respectively, $v_{q\bar{q}} = \sqrt{(k_1.k_2)^2-m_q^4}{\ \big /}(E_1E_2)$ is the relative velocity of the incoming $q\bar{q}$ pair, and $\sigma_{q\bar{q}}^{l^{-}l^{+}}$ is the cross section for the  production of the dilepton from a quark-antiquark pair \cite{Kajantie:1986dh, Ruuskanen:1989tp}. We neglect the rest masses of both quarks and leptons, and as a result the leading order cross section for producing a dilepton of mass $M$ is
\be 
\sigma_{q\bar{q}}^{l^{-}l^{+}} = \frac{4\pi}{3}N_c(2s+1)^2 \frac{\alpha^2}{M^2} \sum_{i=1}^{N_f} e_i^2 \ , 
\ee
which becomes $\sigma =  80\pi \alpha^2 {\big /} 9M^2$ when considering only the $u$ and $d$ flavors. Also, one has $v_{q\bar{q}} = \displaystyle \frac{M^2}{E_1E_2} = \frac{M^2}{|\bf{k}_1||\bf{k}_2|}$ with the assumption of massless fermions.
For the quarks and anti-quarks, we consider the same anisotropic LRF distributions which we parametrize by an ellipsoidal deformation of the isotropic distribution \cite{Romatschke:2004jh} as
\be
f_q\left({\bf k}\right)=f_{\bar{q}}({\bf k})= f\left({\bf k}\right)= f_{\rm{iso}} \left(\frac{1}{\lambda} \sqrt{\frac{k_x^2}{{\alpha}_x^2} + \frac{k_y^2}{{\alpha}_y^2} +\frac{k_z^2}{{\alpha}_z^2} }\right), \label{ellipsoidalf}
\ee 
where $\lambda$ is a temperature-like scale, and the $\alpha_i$ parameters determine the shape and strength of the ellipsoidal anisotropic deformation. In the framework of 3+1d aHydro, $\lambda$ and $\alpha_i$'s all depend on space-time. For fermions, we use the Fermi-Dirac distribution $f_{{\rm FD}}(k)=1/\left(1+e^k\right) $ for $f_{{\rm iso}}$. 

One can change the parameters as
\ba
{\Lambda} &=&{\lambda}{\alpha_y},\\
{\xi}_1 &=&\left(\frac{\alpha_y}{\alpha_z}\right)^2 -1, \\ 
{\xi}_2 &=&\left(\frac{\alpha_y}{\alpha_x}\right)^2 -1,
\ea
to obtain an anisotropic distribution of the form
\be 
f\left({\bf k}\right)= f_{\rm{iso}} \left(\frac{|{\bf k}|}{\Lambda} \sqrt{1+ \xi_1 (\hat{\bf n}_1 \cdot \hat{\bf k})^2 + \xi_2 (\hat{\bf n}_2 \cdot \hat{\bf k})^2 }\right).
\ee

From now on, we rescale all momenta and masses in the calculation by $\Lambda$, and we set $\Lambda \rightarrow 1$ in the equations. Using the delta function, one can integrate \eqref{rate00} over ${\bf k}_2$ to obtain
\ba
\frac{dN}{d^4x d^4P} &=& \frac{80\pi \alpha^2}{9} \int \frac{d^3{\bf k}}{(2\pi)^6} \frac{1}{E_1 E_2} f(\vec{\bf k}) f(\vec{\bf p} - \vec{\bf k}) \ \delta\left(E_1+E_2-E\right) \\
&=& \frac{10 \alpha^2}{72\pi^5} \int_0^{2\pi}d\phi \int_{-1}^1 dx \int_0^E dk\frac{k}{E-k} f(\vec{\bf k}) f(\vec{\bf p} - \vec{\bf k}) \ \delta\left(E_1+E_2-E\right),
\ea
where $x=\cos \theta$, with $\hat{\bf k}=(\sin \theta \cos \phi,\ \sin \theta \sin \phi,\ \cos \theta)$.

Working in a coordinate system where ${\vec{\bf p}}=(0,\ 0,\ p)$, the last delta function gives 
\be 
k+\sqrt{k^2 + p^2 -2kpx}=E. \label{deltaArg}
\ee  
Using this, the rate can be written as
\ba 
\frac{dN}{d^4x d^4P} &=& \frac{10 \alpha^2}{72\pi^5} \int_0^{2\pi}d\phi \int_{-1}^1 dx \  \frac{k_r}{E-px} \label{anisorate} \\ 
&\times & f\left(k_r\sqrt{1+\xi_1 (\hat{\bf n}_1 \cdot \hat{\bf k}_1)^2 + \xi_2 (\hat{\bf n}_2 \cdot \hat{\bf k}_1)^2 }\right) \nonumber \\ 
&\times & f\left( (E-k_r)\ \sqrt{1+\xi_1 (\hat{\bf n}_1 \cdot \hat{\bf k}_2)^2 + \xi_2 (\hat{\bf n}_2 \cdot \hat{\bf k}_2)^2 }\right), \nonumber
\ea
where $k_r = \displaystyle \frac{M^2}{2(E-px)}$ is the solution of \eqref{deltaArg}. The values of $\xi_1$ and $\xi_2$ determine the strength of the ellipsoidal deformation along $\hat{\bf n}_1$ and $\hat{\bf n}_2$. One must note that in the coordinate system used for hydrodynamic evolution, the  components of dilepton momentum, $\hat{\bf n}_1$ and $\hat{\bf n}_2$ are defined as $\vec{\bf p}=p \left(\sin \theta_p \cos \phi_p,\ \sin \theta_p \sin \phi_p,\ \cos \theta_p \right)$, $\hat{\bf n}_1=(0,\ 0,\ 1)$, and $\hat{\bf n}_2=(1,\ 0,\ 0)$ respectively. However, since we use a rotated coordinate system $(x,\phi)$ for the integration of \eqref{anisorate} where ${\vec{\bf p}}=(0,\ 0,\ p)$, we need to use the expressions for $\hat{\bf n}_i$ and $\hat{\bf k}_j$ in terms of these new coordinates as
\ba 
\hat{\bf n}_1 &=& \left( 0,\ -\sin \theta_p,\ \cos \theta_p \right), \\
\hat{\bf n}_2 &=& \left( \sin \phi_p,\ \cos \theta_p \cos \phi_p,\ \sin \theta_p \cos \phi_p \right), \\
\hat{\bf k}_1 &=& \left( \sqrt{1-x^2}\cos \phi,\ \sqrt{1-x^2}\sin \phi,\ x \right), \\
\hat{\bf k}_2 &=& \frac{1}{E - k_r(x)}\left( -k_r(x) \sqrt{1-x^2}\cos \phi,\ -k_r(x) \sqrt{1-x^2}\sin \phi,\ p -xk_r(x) \right).
\ea 
Setting $\xi_1=\xi_2=0$ in the integration \eqref{anisorate} results in the previously known \cite{Cleymans:1986na, Gorenstein:1989ks, Dumitru:1993vz} isotropic rate 
\be
\frac{dN^{({\rm iso})}}{d^4x d^4P} = \frac{10 \alpha^2}{36\pi^4} \frac{2}{p(e^E-1)} \log \left[ \cosh \left( \frac{E+p}{4}\right)  \sech \left( \frac{E-p}{4} \right) \right], \label{isorate}
\ee
which will be used later for comparisons. For non-zero values of anisotropy parameters $\xi_1$ and $\xi_2$, in general, we perform numerical integration of \eqref{anisorate} to determine the LRF dilepton rate.

 We note that for the limiting case of small anisotropy ($\xi_1, \xi_2 \rightarrow 0$), one can use a Taylor series expansion of the anisotropic distributions $f(\vec{\bf k})$ around the isotropic distribution $f_{\rm iso}(k)$ and write
\be
f(\vec{\bf k}) \approx  f_{\rm iso}(k) - \frac{k}{2} f_{\rm iso}(k) \left[ 1- f_{\rm iso}(k) \right] \left[ \xi_1 (\hat{\bf n}_1 \cdot \hat{\bf k})^2 + \xi_2 (\hat{\bf n}_2 \cdot \hat{\bf k})^2  \right].\label{linearized}
\ee
Using this approximation for the anisotropic distributions in \eqref{anisorate}, an additive correction to the isotropic rate \eqref{isorate} is obtained which is essentially very similar to the viscous corrections to the dilepton rate \cite{Vujanovic:2013jpa}. In viscous hydrodynamics, one typically linearizes around isotropic equilibrium distribution and LRF anisotropies are encoded in the $\delta f$ corrections. The correction $\delta f$ is related to the linearized correction shown in  \eqref{linearized}.  However, in general, the values of the anisotropy parameters $\xi_1$ and $\xi_2$ cannot be assumed to be small, and the Taylor expansion around the isotropic limit will not be valid, even resulting in negative values for the distribution function in some regions of phase space. Also, when the parameters $\xi_1$ and $\xi_2$ are not small, the series expansion in these parameters becomes divergent. Therefore, if one needs to use approximate results for the anisotropic rate, methods based on interpolation might be more reliable than methods based on perturbative expansion. In this paper, we simply integrate the dilepton rate \eqref{anisorate} numerically, and without linearization, using adaptive quadrature. 

\begin{figure}[h]
\centerline{
\includegraphics[width=1.0\linewidth]{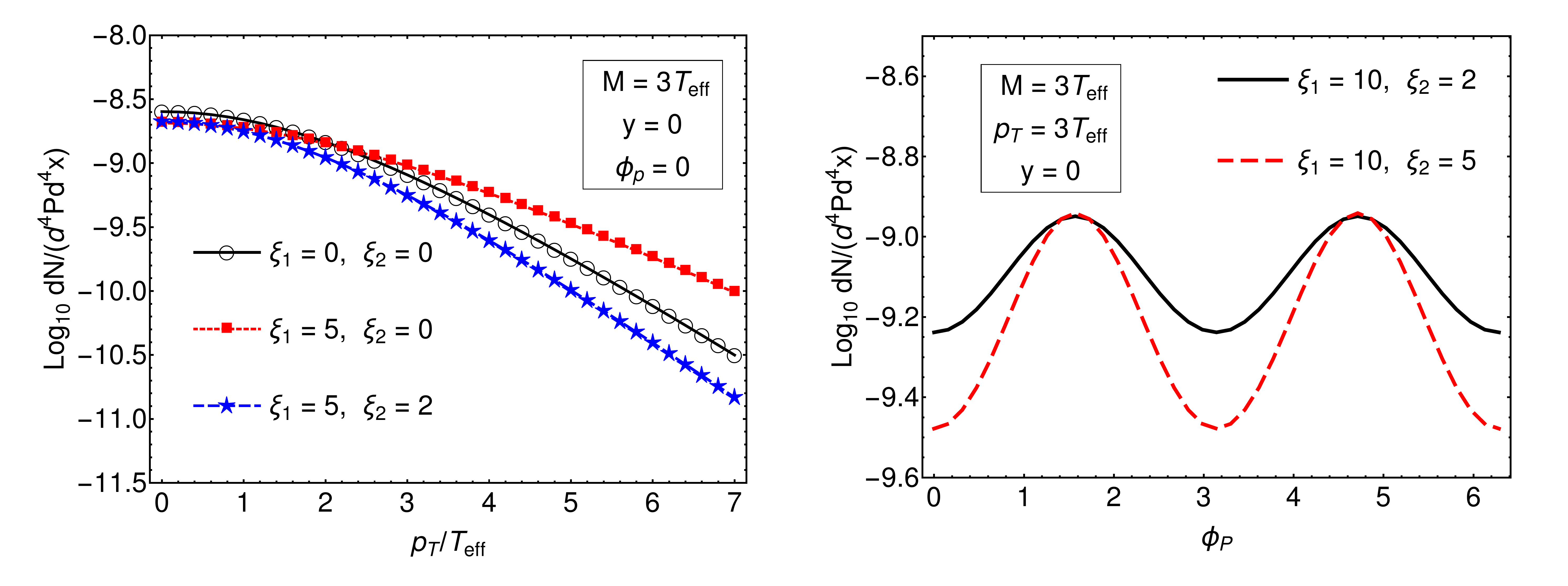}
}
\caption{(Left panel): Differential dilepton production rate in the local rest frame of a QGP fluid element with effective temperature of $T_{\rm eff}$, compared for three cases: isotropic ($\xi_1=\xi_2=0$), spheroidally anisotropic ($\xi_1=5, \ \xi_2=0$), and ellipsoidally anisotropic ($\xi_1=5, \ \xi_2=2$) QGP. \\ 
(Right panel): Differential dilepton production rate as a function of $\phi_P$ for two ellipsoidally anisotropic cases.} 
\label{plot:lrf}
\end{figure}

In the left panel of Fig.~\ref{plot:lrf}, the local rest frame dilepton rate \eqref{anisorate} as a function of transverse momentum is compared for the three cases of  an isotropic, spheroidally anisotropic $(\xi_2=0)$, and ellipsoidally anisotropic QGP with an equivalent effective temperature. The effective temperature is determined by the condition of equal energy densities calculated from the anisotropic and isotropic momentum distributions, i.e. $\epsilon(\Lambda,\xi_1,\xi_2)= \epsilon_{\rm iso}(T_{\rm eff})$, which leads to
\be 
\left(\frac{T_{\rm eff}}{\Lambda(\xi_1, \xi_2)}\right)^4 = \frac{1}{4\pi} \int_0^{2\pi}d\phi \int_{-1}^1 \frac{d(\cos\theta)}{\left(1+\xi_1 \cos^2\theta + \xi_2 \sin^2\theta \cos^2\phi\right)^2}\ .
\ee

From the left panel of Fig.~\ref{plot:lrf}, it can be seen that a small transverse plane anisotropy ($\xi_2 \neq 0$) can induce a significant difference with the spheroidally anisotropic rate at higher dilepton transverse momenta. In addition, contrary to the isotropic and spheroidally anisotropic cases, the rate for the ellipsoidally anisotropic case depends on the azimuthal direction of dilepton momentum. This $\phi_P$-dependence is shown in the right panel of Fig.~\ref{plot:lrf}. One should note that, after integrating over the space-time and dilepton mass/momentum, considering boost transformation, the manner in which the differences in the distribution functions can affect the experimental observables becomes non-trivial and complicated. 

\subsection{Space-time integrated dilepton yield and flow}

In order to calculate the QGP dilepton yield and flow coefficients, one needs to convolve the differential LRF production rate \eqref{anisorate} together with the space-time evolution of the strongly interacting medium. The parameters like $\eta/s$ control the aHydro evolution which then provides the full 3+1d evolution of the local temperature-like scale $\lambda(x)$, the local anisotropy parameters $\alpha_i(x)$ (or equivalently $\xi_i(x)$) and fluid velocity entering the anisotropic distribution functions used in the calculation of dilepton rates. In this way, dissipative corrections due to the shear viscosity are automatically included in both the aHydro evolution and anisotropic dilepton production rates.  Dileptons with four-momentum $P'^{\mu}$ emitted in the LRF of a fluid element with four-velocity $u^{\mu} = \gamma \left(1,\ v_x,\ v_y,\ v_z \right)$ get boosted to the lab frame four-momentum $P^{\mu}=\Lambda_{\mu}^{\nu}(u) P'^{\nu}$ with Lorentz boost $\Lambda_{\mu}^{\nu}(u)$ and Lorentz factor $\gamma = 1/\sqrt{1-v^2}$. The four variables of dilepton mass $M$, transverse momentum $p_T$, momentum rapidity $y$, and momentum azimuthal angle $\phi_p$ are used to characterize the four-momentum of the detected dilepton as
$
P^{\mu} = \left(  m_T \cosh y,\     p_T \cos \phi_p,\   p_T \sin \phi_p,\  m_T \sinh y  \right),
$  
where $m_T = \sqrt{M^2+p_T^2}$ is the transverse mass. Integrating over space-time, the differential yield of dileptons becomes
\be
\frac{dN}{MdMp_Tdp_Tdyd\phi_p} = \int d^4x \ \frac{dN}{d^4xd^4P }\ \label{int4x}.
\ee 
To calculate the mass/transverse momentum dependence of the yield and the flow, one usually integrates over  the $p_T$ or $M$ variables to obtain
\ba
\frac{dN}{MdMdyd\phi_p} &=&   \int_{p_T^{\rm min}}^{p_T^{\rm max}} p_T dp_T  \frac{dN}{MdMp_Tdp_Tdyd\phi_p}\ ,    \label{intpt}\\
\frac{dN}{p_Tdp_Tdyd\phi_p} &=&  \int_{M^{\rm min}}^{M^{\rm max}}MdM  \frac{dN}{MdMp_Tdp_Tdyd\phi_p}\ . \label{intm}
\ea
In the reaction plane and at fixed rapidity, the $M$ and $p_T$ dependent flow coefficients $v_n$ are defined using expansions in terms of $\cos n\phi_p$ functions
\ba
\frac{dN}{MdMdyd\phi_p} &=&  \frac{1}{2\pi} \frac{dN}{MdMdy} \Big[ 1+ 2v_1\left(M\right) \cos \phi_p  + 2v_2\left(M\right) \cos 2\phi_p  + \cdots  \Big]  , \\
\frac{dN}{p_Tdp_Tdyd\phi_p} &=&  \frac{1}{2\pi} \frac{dN}{p_Tdp_Tdy} \Big[ 1+ 2v_1\left(p_T\right) \cos \phi_p  + 2v_2\left(p_T\right) \cos 2\phi_p  + \cdots  \Big]. 
\ea
   
\subsection{Parameters and settings} \label{settings}

For the space-time evolution of the QGP, we use 3+1d aHydro model with parameters set for Pb-Pb collisions at $\sqrt{s}=2.76$\ TeV at the LHC. The free parameters of this hydrodynamic model, such as the shear viscosity to entropy density ratio and the initial central temperature, are set based on the best fits of the model calculations to the soft hadron spectra \cite{Alqahtani:2017tnq}. The initial state is modeled using a smooth Glauber model \cite{Miller:2007ri} and the centrality classes are represented by their mean impact parameter value. The initial proper time is taken to be $\tau_0 = 0.25\ {\rm fm}/c$. The value of shear viscosity to entropy density ratio $\eta/s$ is assumed to be constant during the aHydro evolution and is taken to be multiples of the $(\eta/s)_{\rm KSS}=1/4\pi$, which is the lower bound suggested by AdS/CFT conjecture \cite{Kovtun:2004de}. The freeze-out temperature is taken to be $0.130$\ GeV, and the initial central temperature for each value of $\eta/s$ is adjusted to obtain the best fits to soft hadron spectra, giving $T_0=$ 0.63, 0.6, and 0.58 GeV for $4\pi\eta/s$ values of 1, 2, and 3 respectively, assuming no initial momentum anisotropy.\footnote{See \citep{Alqahtani:2017tnq} for more aHydro calculations of soft hadron spectra assuming no initial momentum anisotropy.} The pair $(4\pi\eta/s,\ T_0)=(2,\ 0.6\ {\rm GeV})$ was found to result in the best fit to the hadron spectra, therefore we use it as our reference setting for the study of dilepton production. For momentum-anisotropic initial conditions, we find that with $(\alpha_x,\ \alpha_y,\ \alpha_z){\big |}_{\tau_0}=(1,\ 1,\ 0.5)$, an initial temperature of 0.58 GeV with $4\pi\eta/s=2$ reproduces the hadron spectra with same accuracy as the case with no initial momentum anisotropy (See Sec.~\ref{inianiso}).
The aHydro evolution uses a quasiparticle equation of state \citep{Alqahtani:2017jwl} extracted from lattice QCD calculations \cite{Borsanyi:2010cj}. The hadronic freeze-out and decays are performed using the THERMINATOR 2 Monte Carlo event generator \cite{Chojnacki:2011hb}.

In all of the calculations presented in this paper, we have considered dileptons with rapidity $y=0$ in the lab frame. The focus of this paper is on dileptons produced from the QGP phase and we present the results only for $M \geq 1$ GeV and $p_T \geq 1$ GeV. To consider only the QGP phase contribution to dilepton emission, the rate from regions of fluid with $T_{\rm effective}<T_c$ is set to 0, where $T_c$ is the critical temperature which is taken to be 0.155 GeV. When integrating over $M$ or $p_T$ according to \eqref{intm} or \eqref{intpt}, we use the integration regions $1 <M<20$ GeV and $1<p_T<20$ GeV. In this paper, we use Monte Carlo integration to calculate \eqref{int4x}, \eqref{intpt}, and \eqref{intm}.

\section{Results and discussion}
\label{results}

In this section, our results  for the mid-rapidity yields and elliptic flow of QGP-produced dileptons generated in Pb-Pb collisions at $\sqrt{s}=2.76$ TeV at the LHC are discussed. We emphasize the importance of incorporating momentum-anisotropic local rest frame dilepton rates along with the relativistic anisotropic hydrodynamical evolution of the QGP. The effects of shear viscosity to entropy density ratio on the production and flow of dileptons are also investigated. We provide predictions for the invariant mass and transverse momentum  dependence of the intermediate mass QGP-sourced dilepton yields and elliptic flow in different centrality classes. We also discuss the effects of initial momentum anisotropy on the results.

\subsection{Effects of local rest frame momentum anisotropy}

\begin{figure}[h]
\centerline{
\includegraphics[width=1.0\linewidth]{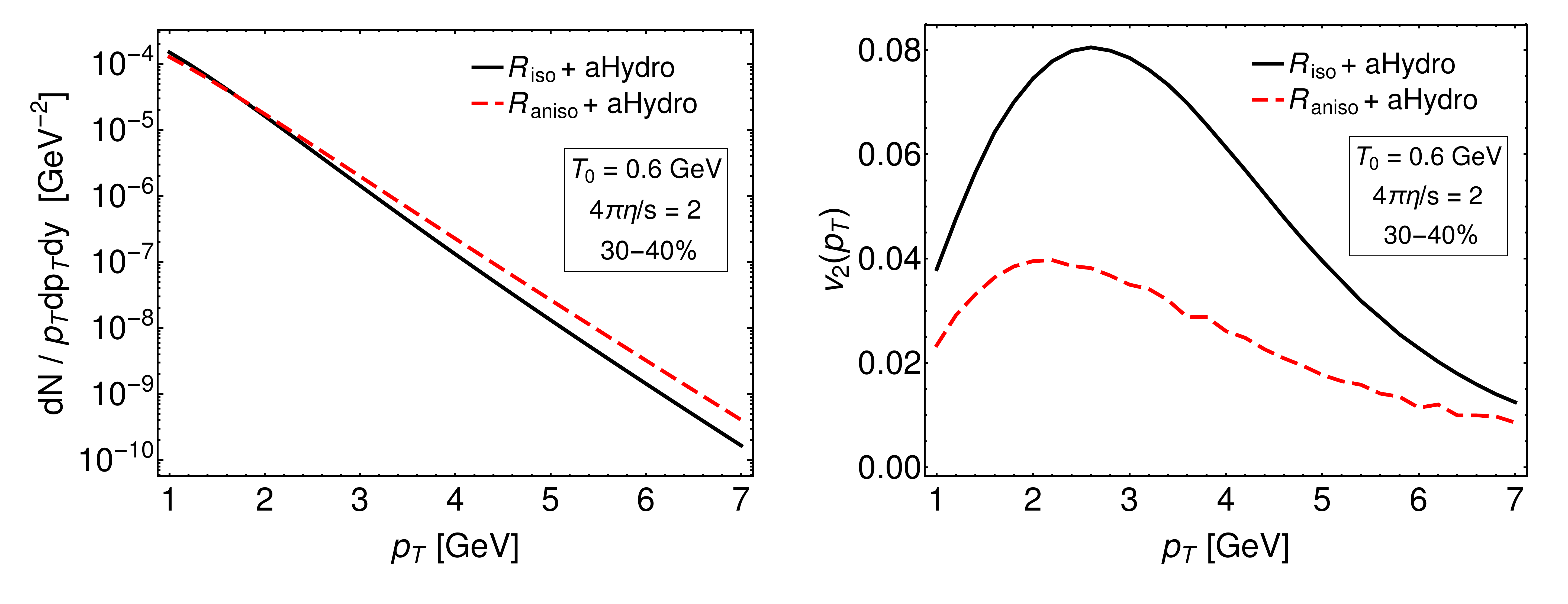}
}
\caption{Thermal dilepton yield (left) and $v_2(p_T)$ (right) in two cases: (\textit{solid line}) aHydro QGP evolution convolved with the isotropic LRF dilepton rate \eqref{isorate}, and (\textit{dashed line}) aHydro QGP evolution convolved with the anisotropic LRF dilepton rate \eqref{anisorate}.} 
\label{plot:isoaniso}
\end{figure}

In order to have a theoretically consistent calculation of dilepton yield and flow within a framework of non-ideal hydrodynamics like aHydro, one needs to include the corresponding non-ideal effects on the local rest frame dilepton rates. However, one might argue that the contribution of LRF non-ideal effects on the final observables might not be comparable to the effects of the overall hydrodynamic evolution. In order to investigate the effects of momentum anisotropy,  in Fig.~\ref{plot:isoaniso}, we compare results of convolving the isotropic rate \eqref{isorate} and the complete anisotropic rate \eqref{anisorate} with the 3+1d aHydro evolution. For the isotropic rate case, we computed the local effective temperature by Landau matching the energy density obtained from the non-equilibrium aHydro evolution \citep{Alqahtani:2017mhy}. 
The comparison is done for 30-40\% centrality class, $4\pi\eta/s=2$, and $T_0=0.6\ {\rm GeV}$. The left panel of Fig.~\ref{plot:isoaniso} shows that inclusion of the anisotropic rate has a visible effect on the dilepton yield and can be seen to harden the spectrum. The right panel of Fig.~\ref{plot:isoaniso} shows that there is significant difference in the two cases for $v_2(p_T)$. The effects of momentum anisotropy in LRF dilepton rate reduce (isotropizes) the dilepton flow generated through coupling to the hydrodynamic evolution. We also mention that we verified explicitly that the same effects do not induce a change in the other harmonics of dilepton flow.   

\begin{figure}[h]
\centerline{
\includegraphics[width=1.0\linewidth]{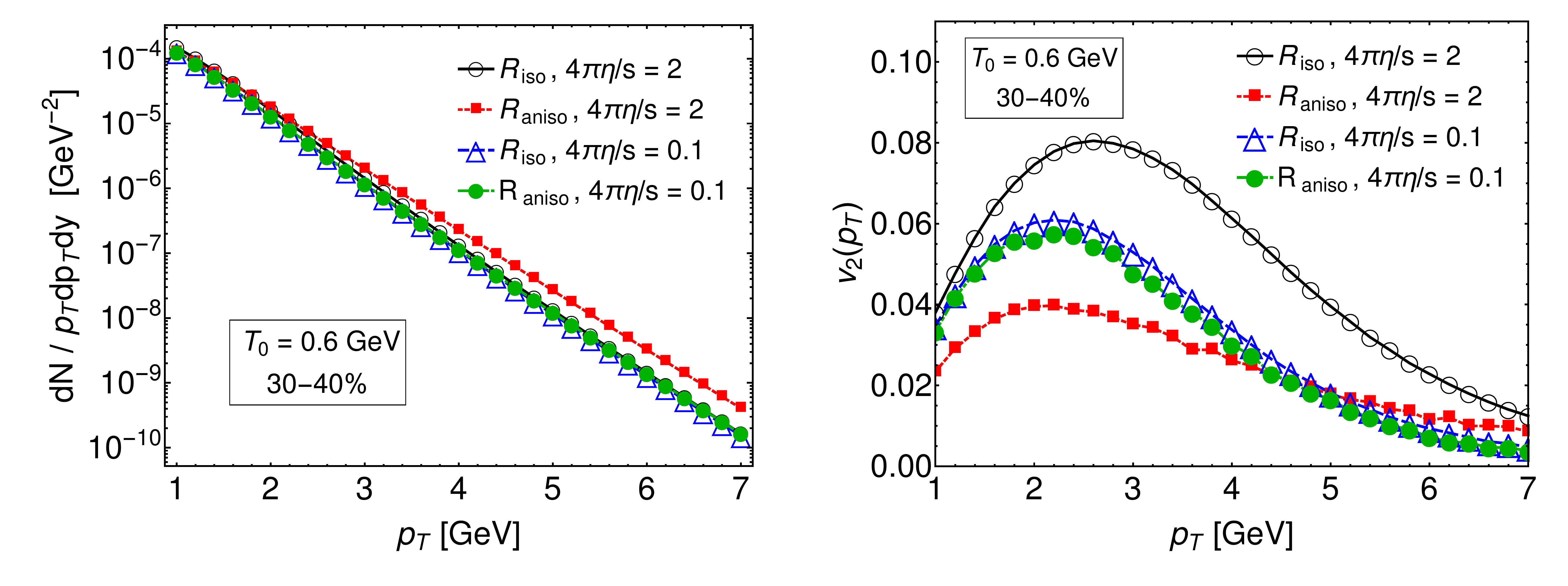}
}
\caption{Thermal dilepton yield (left) and $v_2(p_T)$ (right) for four cases: (\textit{empty circle markers}) aHydro QGP evolution (with $4\pi\eta/s=2$) convolved with isotropic LRF dilepton rate, (\textit{filled red square marker}) aHydro QGP evolution (with $4\pi\eta/s=2$) convolved with anisotropic LRF dilepton rate, (\textit{empty blue triangle marker}) aHydro QGP evolution (near ideal with $4\pi\eta/s=0.1$) convolved with isotropic LRF dilepton rate, and (\textit{filled green circle marker}) aHydro QGP evolution (near ideal with $4\pi\eta/s=0.1$) convolved with anisotropic LRF dilepton rate.} 
\label{plot:isoaniso_smallE}
\end{figure}

In Fig.~\ref{plot:isoaniso_smallE} we also compare with results obtained by setting $4\pi\eta/s=0.1$ which causes the hydrodynamic evolution to be closer to the ideal case. For near-ideal fluid, as one expects, the results from isotropic and anisotropic LRF rates are close. Dilepton elliptic flow from aHydro evolution convolved with isotropic LRF rate is larger than the near-ideal case. However, inclusion of momentum-anisotropic corrections to the LRF dilepton rate decreases the magnitude of dilepton elliptic flow and makes it even smaller than the near-ideal case. This comparison of the near-ideal and the $4\pi\eta/s=2$ cases, based on Fig.~\ref{plot:isoaniso_smallE}, can only be taken as a qualitative measure since the initial temperatures are set to the same value of 0.6 GeV in both cases and not tuned for the $4\pi\eta/s=0.1$ case using soft hadron spectra.

\subsection{Effects of shear viscosity-entropy density ratio}

\begin{figure}[h]
\centerline{
\includegraphics[width=1.0\linewidth]{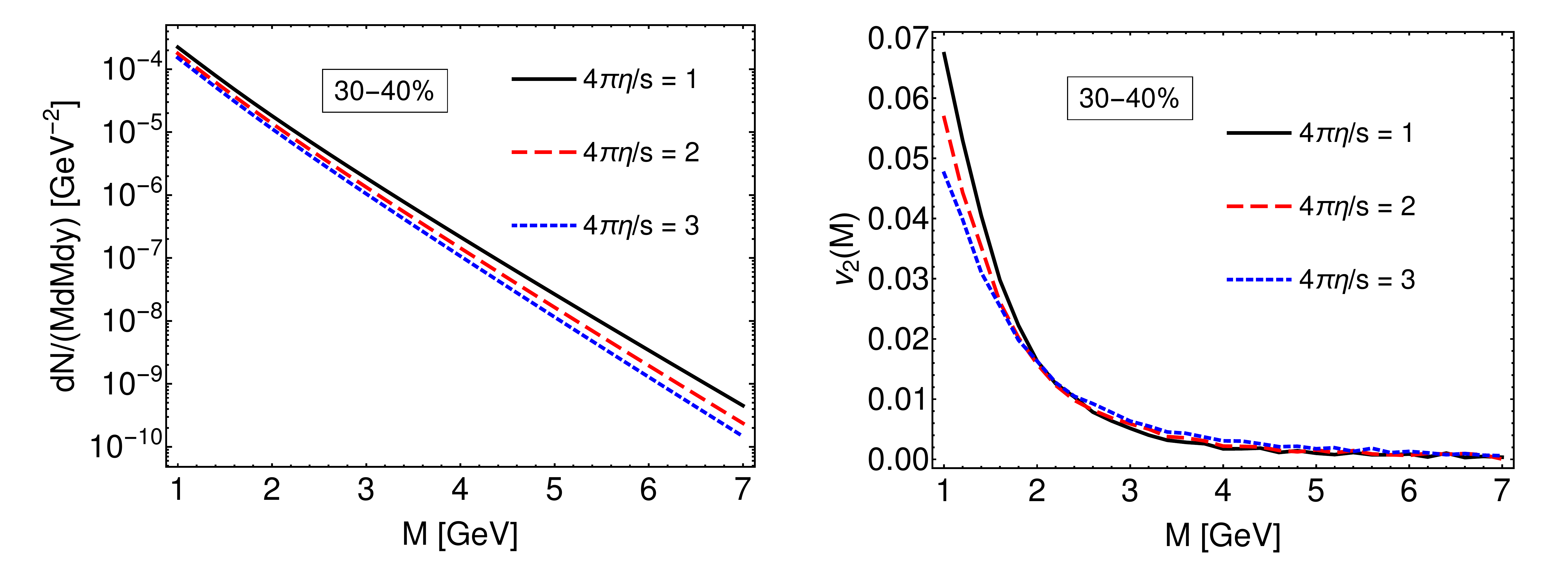}
}
\caption{Invariant mass dependence of the mid-rapidity ($y=0$) thermal dilepton yield (left) and $v_2$ (right) for different $\eta/s$ and 30-40\% centrality class, assuming initial momentum isotropy. The initial temperature in each case is set by best fits of aHydro results to hadronic spectra which results in $T_0=\{0.63,\ 0.6,\ 0.58\} \ {\rm GeV}$ for $4\pi\eta/s=\{1,\ 2,\ 3\}$, respectively.} 
\label{plot:etabar_123_M}
\end{figure}

\begin{figure}[h]
\centerline{
\includegraphics[width=1.0\linewidth]{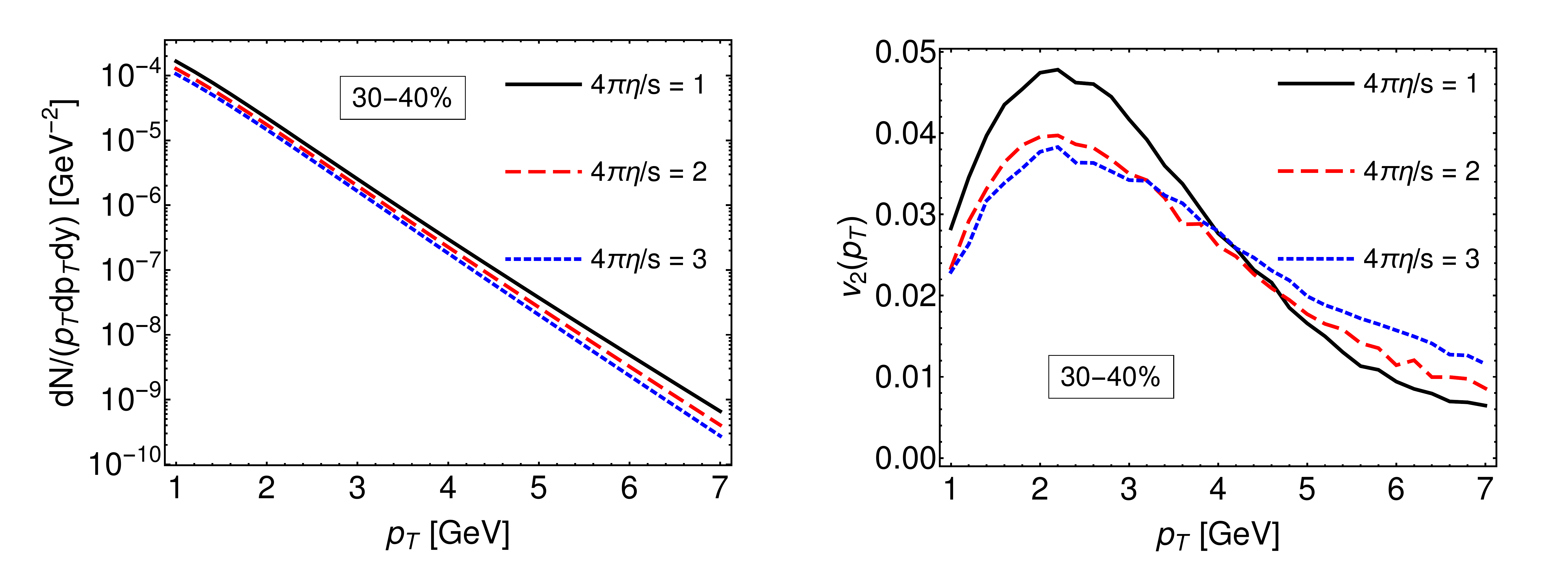}
}
\caption{Transverse momentum dependence of mid-rapidity ($y=0$) thermal dilepton yield (left) and $v_2$ (right) for different $\eta/s$ and 30-40\% centrality class, assuming initial momentum isotropy. Initial temperatures used were the same as in Fig.~\ref{plot:etabar_123_M}.} 
\label{plot:etabar_123_pT}
\end{figure}

We compare the yield and flow of thermal dileptons for three cases with $4\pi\eta/s$ values of 1, 2, and 3, all set to their corresponding fit initial temperature values (see Sec.~\ref{settings}). From Fig.~\ref{plot:etabar_123_M} we see that the $M$-dependence of $v_2$ is sensitive to $\eta/s$ only when $M \lesssim 2$ GeV. However, in Fig.~\ref{plot:etabar_123_pT} a non-trivial behavior of $p_T$-dependent results can be clearly seen in which for the $p_T \lesssim 4\ {\rm GeV}$ region, the smaller $\eta/s$ result in larger $v_2$ but, for $p_T \gtrsim 4\ {\rm GeV}$, the order becomes reversed i.e. the larger $\eta/s$ corresponds to higher values of $v_2$.  

\subsection{Centrality dependence of dilepton yield and flow}

\begin{figure}[h]
\centerline{
\includegraphics[width=1.0\linewidth]{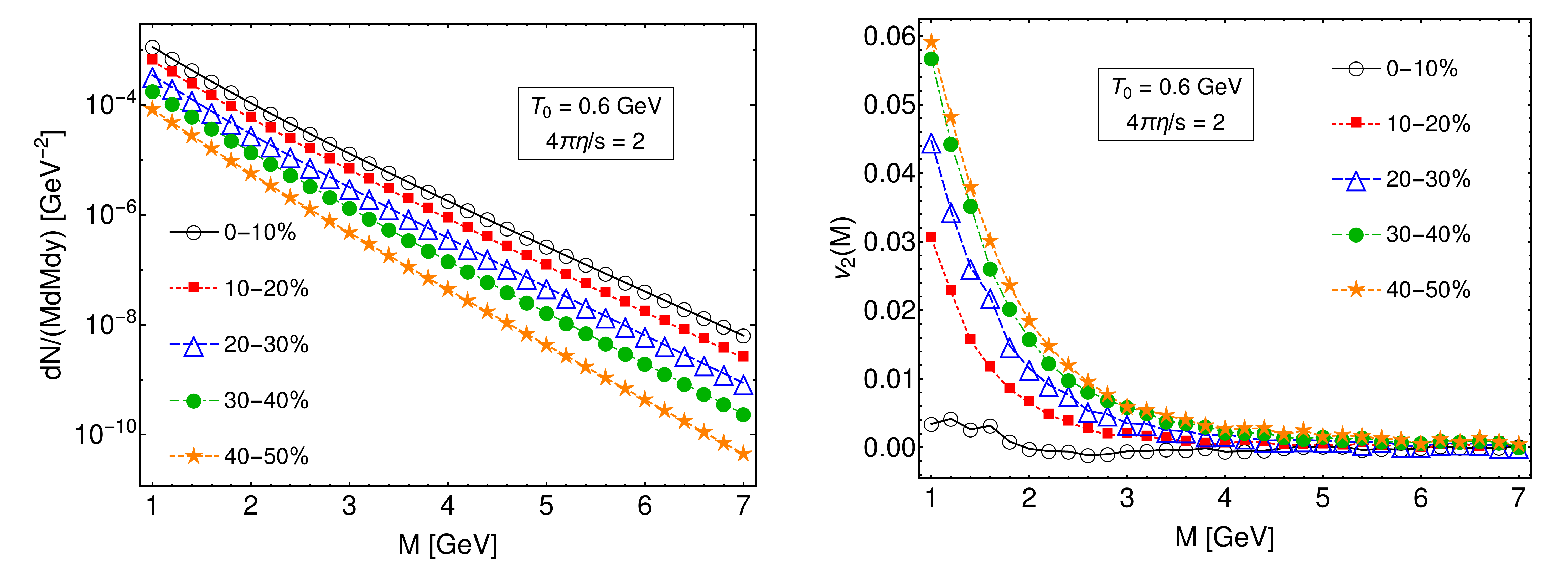}
}
\caption{Invariant mass dependence of mid-rapidity ($y=0$) thermal dilepton yield (left) and $v_2$ (right) for different centrality classes, assuming initial momentum isotropy, $4\pi\eta/s=2$, and initial central temperature of $T_0=$ 0.6 GeV.} 
\label{plot:Centrality_M_2_600}
\end{figure}

\begin{figure}[h]
\centerline{
\includegraphics[width=1.0\linewidth]{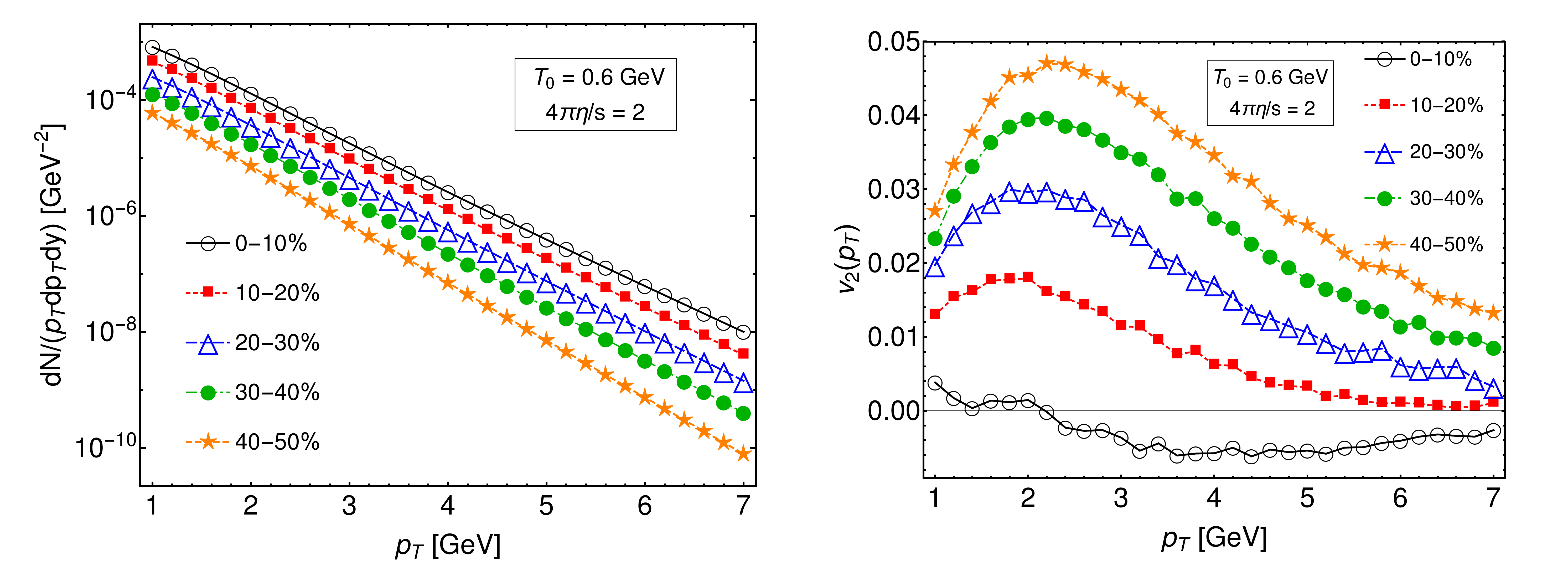}
}
\caption{Transverse momentum dependence of mid-rapidity ($y=0$) thermal dilepton  yield (left) and $v_2$ (right) for different centrality classes, assuming initial momentum isotropy, $4\pi\eta/s=2$, and initial central temperature of $T_0=$ 0.6 GeV.} 
\label{plot:Centrality_pT_2_600}
\end{figure}

\begin{figure}[h]
\centerline{
\includegraphics[width=1.0\linewidth]{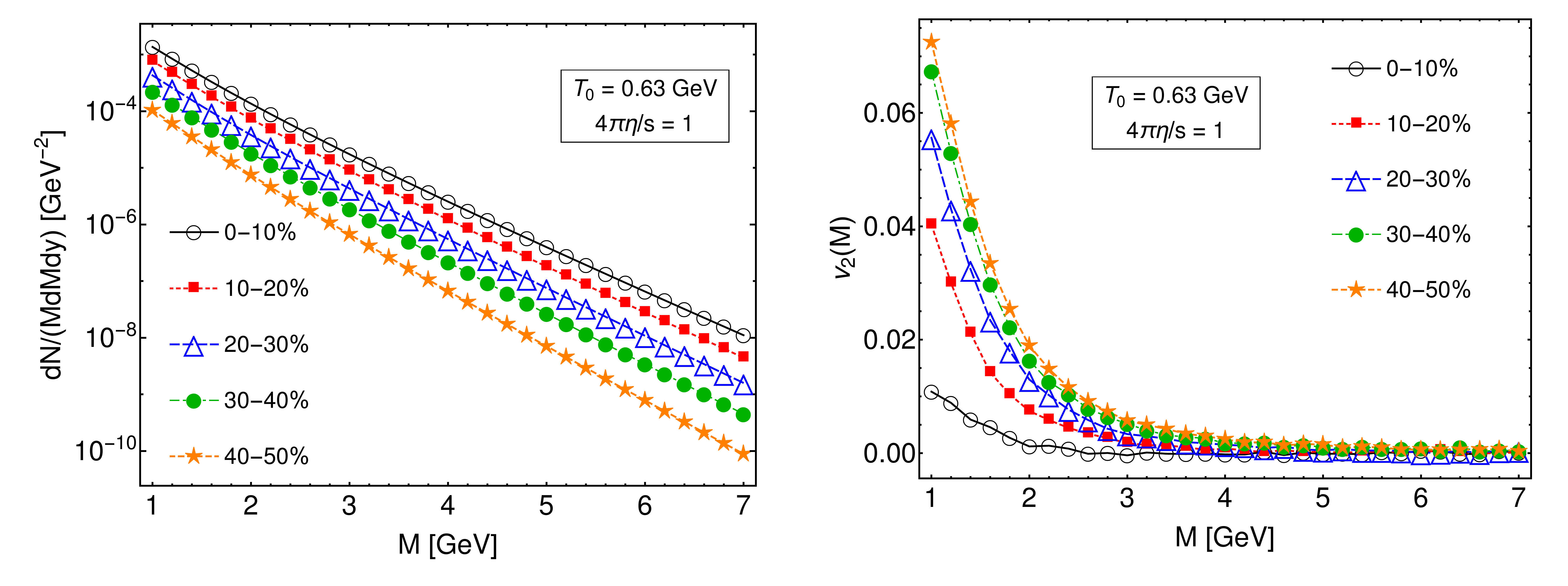}
}
\caption{Invariant mass dependence of mid-rapidity ($y=0$) thermal dilepton yield (left) and $v_2$ (right) for different centrality classes, assuming initial momentum isotropy, $4\pi\eta/s=1$, and initial central temperature of $T_0=$ 0.63 GeV.} 
\label{plot:Centrality_M_1_630}
\end{figure}

\begin{figure}[h]
\centerline{
\includegraphics[width=1.0\linewidth]{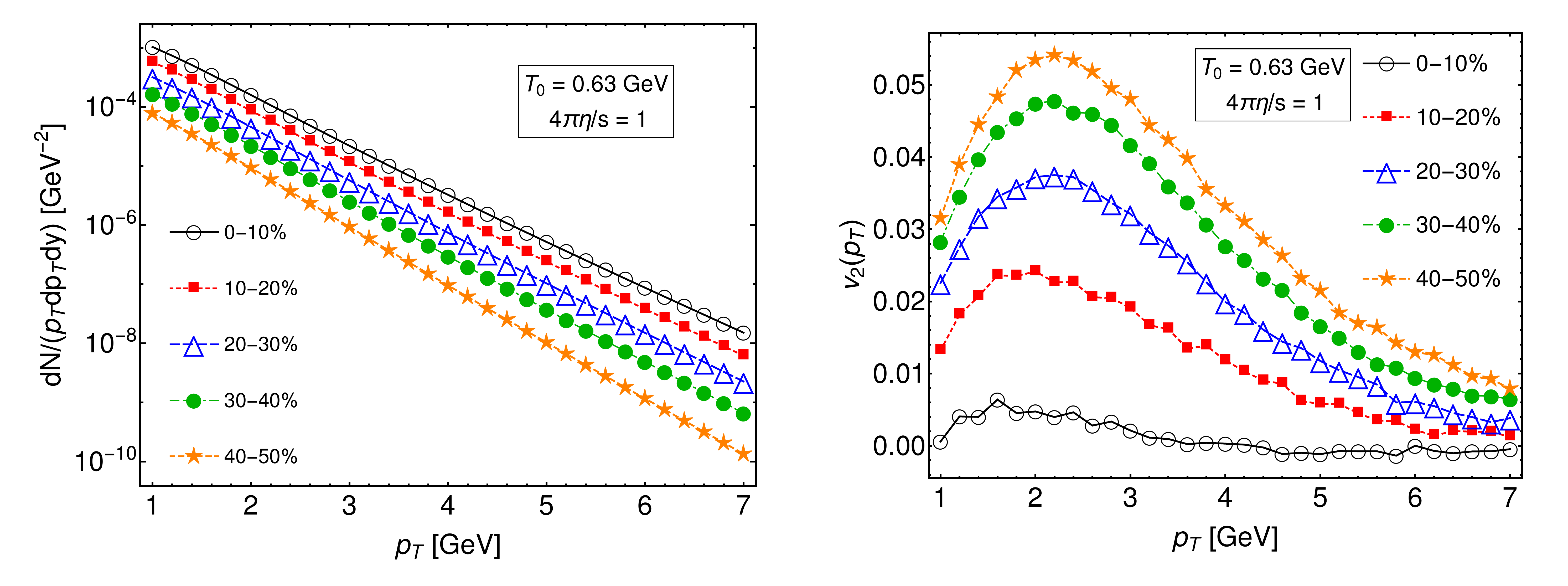}
}
\caption{Transverse momentum dependence of mid-rapidity ($y=0$) thermal dilepton yield (left) and $v_2$ (right) for different centrality classes, assuming initial momentum isotropy, $4\pi\eta/s=1$, and initial central temperature of $T_0=$ 0.63 GeV.} 
\label{plot:Centrality_pT_1_630}
\end{figure}

The primary source of the anisotropic flow of particles produced in heavy-ion collisions is believed to be the spatial asymmetry of the initial state \cite{Heinz:2013th}. To investigate the dependence of the thermal dilepton emission on the centrality of the collision, we have calculated the differential yields and elliptic flow coefficient for the LHC collisions with mean impact parameters of the 0-10, 10-20, 20-30, 30-40, and 40-50 percent centrality classes. The results for $T_0=0.6$ GeV and $4\pi \eta/s=2$ are shown in Fig.~\ref{plot:Centrality_M_2_600} as a function of $M$, and in Fig.~\ref{plot:Centrality_pT_2_600} as a function of $p_T$. As expected, for the more central collisions the production yields are higher and elliptic flow is smaller.  
One interesting finding shown in Fig.~\ref{plot:Centrality_pT_2_600} is that for central collisions, the calculated elliptic flow coefficient of the QGP dileptons appears to have small but negative values for $p_T \gtrsim 2$ GeV. However, this result could be merely due to the numerical uncertainties. We also note that for a more reliable interpretation of the results for central collisions, one must include the effects of fluctuating initial conditions in the model which are not included in this paper. The same calculations for $4\pi \eta/s=1$ (with adjusted initial central temperature $T_0=0.63\ {\rm GeV}$) are presented in Fig.~\ref{plot:Centrality_M_1_630} and Fig.~\ref{plot:Centrality_pT_1_630}. The negative flow coefficients for central collisions are not apparent in this case. Negative values of $v_2$ for photons, protons, and $J/\psi$'s have been discussed in other papers \cite{Danielewicz:1998vz, Pinkenburg:1999ya, Turbide:2005bz, Krieg:2007bc}. 

\begin{figure}[h]
\centerline{
\includegraphics[width=1.0\linewidth]{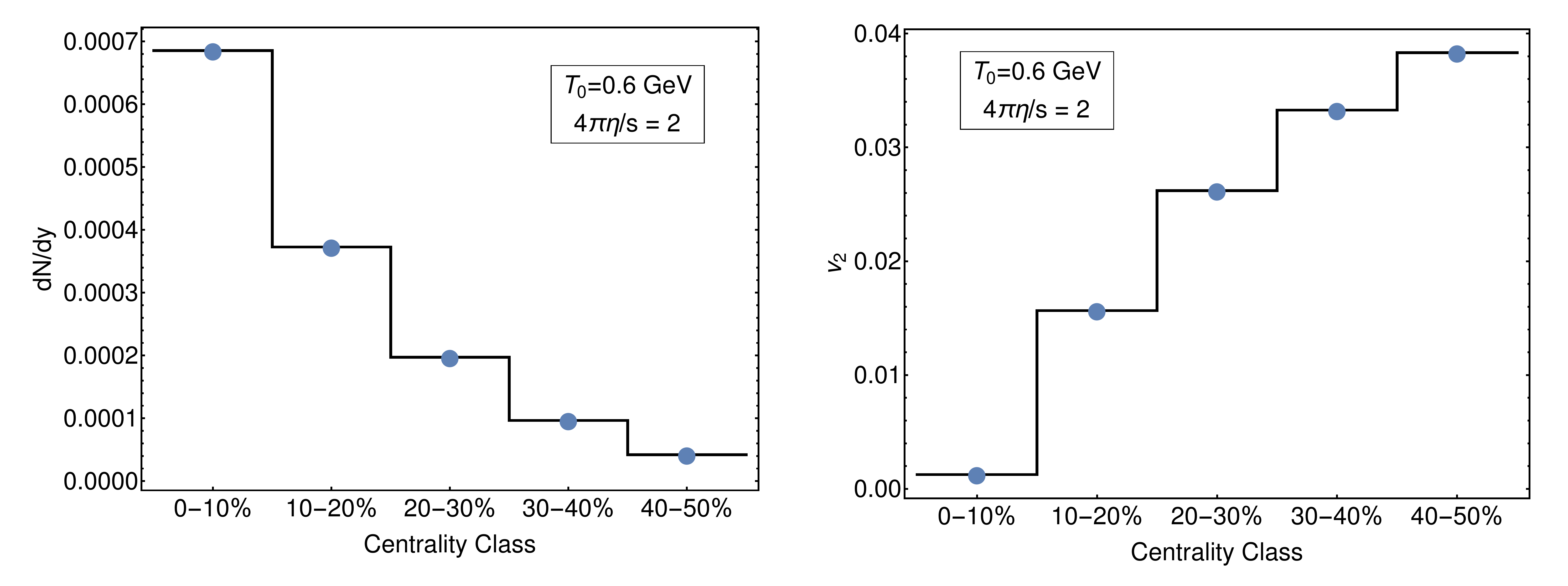}
}
\caption{Centrality dependence of ($M,p_T$)-integrated mid-rapidity ($y=0$) thermal dilepton yield (left) and $v_2$ (right), assuming initial momentum isotropy, $4\pi\eta/s=2$, and initial central temperature of $T_0= 0.6 {\ \rm GeV}$.} 
\label{plot:integrated_centrality}
\end{figure}

The predictions for integrated (over $M$ and $p_T$) mid-rapidity yield and elliptic flow of thermal dileptons for different centrality classes are shown in Fig.~\ref{plot:integrated_centrality}, where $4\pi \eta/s=2$ and $T_0=0.6$ GeV were used in the calculation. 

\subsection{Effects of initial momentum anisotropy} \label{inianiso}

All of the results in previous sections of this paper were calculated with the assumption of momentum isotropy for the initial state of the hydrodynamical evolution. It would be interesting to see whether the dilepton emission is affected by initial momentum anisotropies at $\tau_0$. The existence of a non-equilibrium attractor in various theories, e.g. kinetic theory and strongly coupled gauge duals \cite{Florkowski:2017olj, Strickland:2017kux} suggests that different initially anisotropic states converge to the same dynamics well before the end of the evolution of system. Higher energy dileptons produced during the initial stages might provide less distorted information about the initial momentum distributions since they escape freely. 

\begin{figure}[h]
\centerline{
\includegraphics[width=1.0\linewidth]{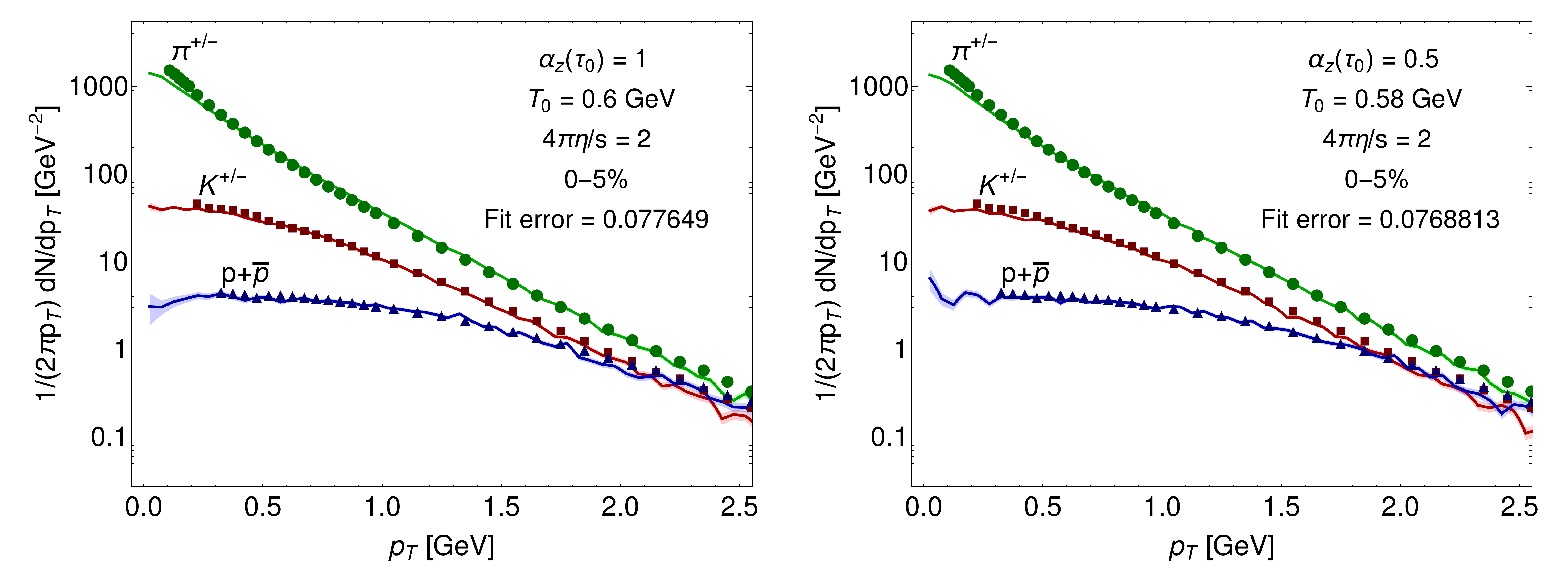}
}
\caption{Best fits of aHydro model calculations to experimental data for $\sqrt{s}=2.76\ {\rm TeV}$ soft hadron spectra $(|y|<1)$ for two settings of the model: (left panel) a QGP with no initial momentum anisotropy ($\alpha_{\{x,y,z\}}(\tau_0)=1$) and (right panel) a QGP with initial momentum anisotropy ($\alpha_{\{x,y\}}(\tau_0)=1,\ \alpha_z(\tau_0)=0.5$). The experimental data shown are from ALICE collaboration \cite{Abelev:2013vea}.}
\label{plot:spectra}
\end{figure}

To investigate the effects of initial momentum anisotropy, we compare the yields and $v_2$ of dileptons for three cases: (1) QGP with initially isotropic momentum distributions ($\alpha_{\{x,y,z\}}(\tau_0)=1$) with $4\pi\eta/s=2$ and $T_0=0.6\ {\rm GeV}$, (2) QGP with initial momentum anisotropy ($\alpha_z(\tau_0)=0.5$) with same $4\pi\eta/s=2$ and $T_0=0.6\ {\rm GeV}$, and (3) QGP with a momentum-anisotropic initial condition ($\alpha_z(\tau_0)=0.5$) with $4\pi\eta/s=2$ but the initial central temperature re-tuned in order to reproduce the experimentally observed final hadronic spectra. Although the value of $T_0=0.6\ {\rm GeV}$ was tuned for  $4\pi\eta/s=2$ assuming no initial momentum anisotropy (the case for which the comparison to the hadron spectra is shown in the left panel of Fig.~\ref{plot:spectra}), for a QGP with initial momentum anisotropy, we found that, with the same value of $\eta/s$, the initial temperature of $T_0=0.58\ {\rm GeV}$ provides the best fit of model calculations to the hadron spectra which is shown in the right panel of Fig.~\ref{plot:spectra}.

\begin{figure}[h]
\centerline{
\includegraphics[width=1.0\linewidth]{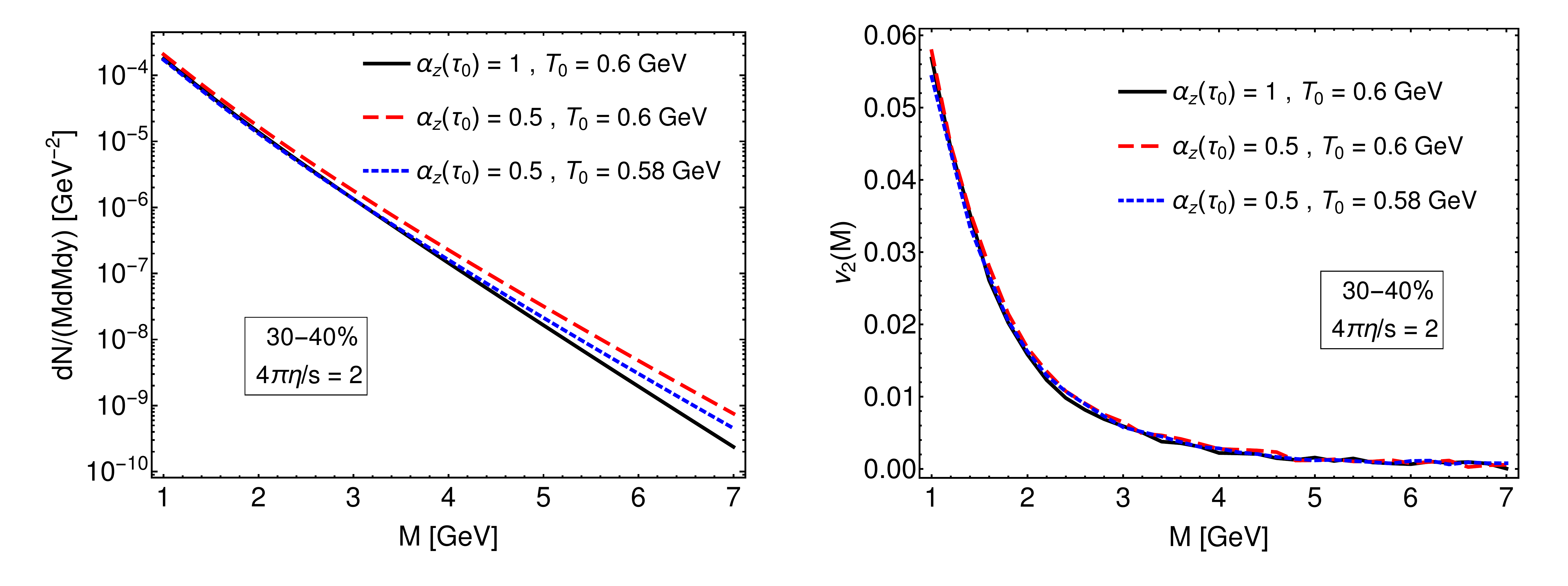}
}
\caption{Invariant mass dependence of mid-rapidity ($y=0$) QGP dilepton yield (left) and $v_2$ (right) for 30-40\% centrality class, compared for three cases: initial momentum isotropy ($\alpha_z(\tau_0)=1$) with $T_0= 0.6 \ {\rm GeV}$, initial spheroidal momentum anisotropy ($\alpha_z(\tau_0)=0.5$) with same $T_0= 0.6 \ {\rm GeV}$, and initial spheroidal momentum anisotropy ($\alpha_z(\tau_0)=0.5$) with adjusted initial central temperature of $T_0= 0.58 \ {\rm GeV}$ based on our best fit to hadronic spectra (See Fig.~\ref{plot:spectra}).}
\label{plot:IniAniso_M}
\end{figure}

\begin{figure}[h]
\centerline{
\includegraphics[width=1.0\linewidth]{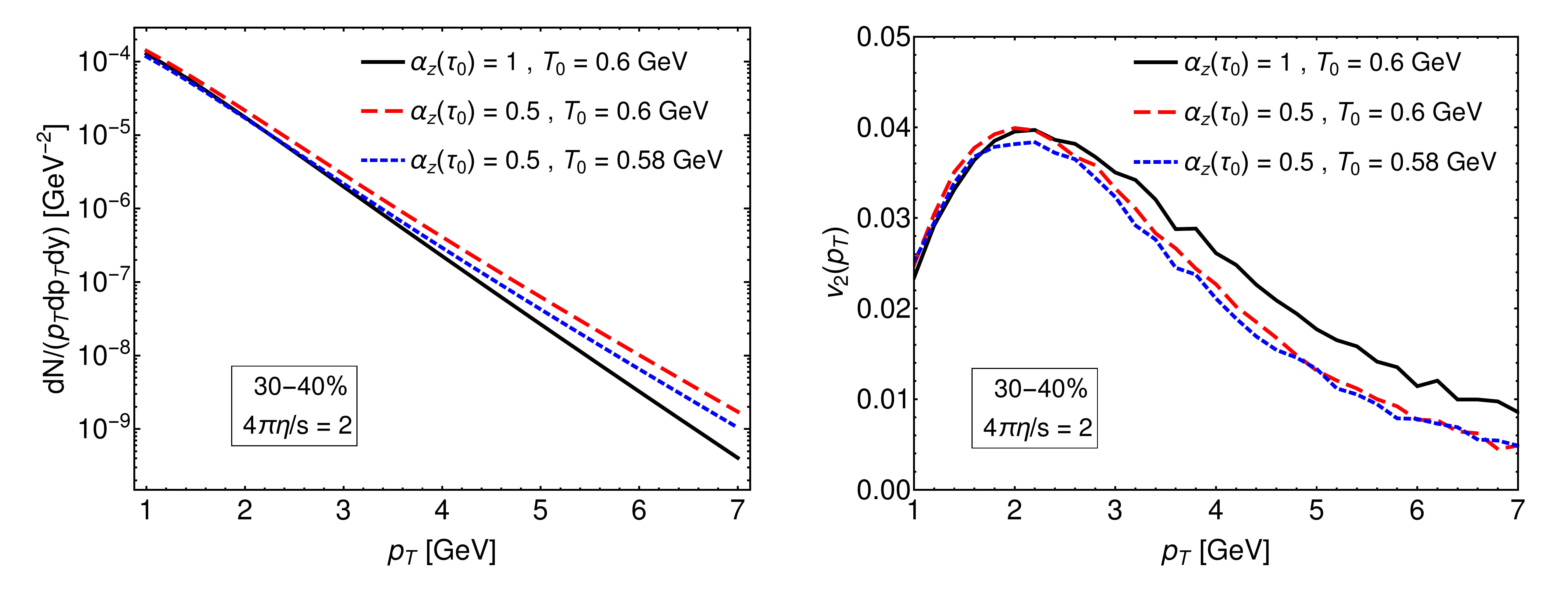}
}
\caption{Transverse momentum dependence of mid-rapidity ($y=0$) QGP dilepton yield (left) and $v_2$ (right) for 30-40\% centrality class, compared in same three cases as Fig.~\ref{plot:IniAniso_M}.}
\label{plot:IniAniso_pT}
\end{figure}

In Fig.~\ref{plot:IniAniso_M} we compare the $M$-dependence of dilepton yields and $v_2$ for the three cases mentioned above. 
The results show that initial momentum anisotropy induces higher yields for higher mass dileptons. However, the $v_2(M)$ can not clearly distinguish between the three cases.  In Fig.~\ref{plot:IniAniso_pT} the same comparison is presented for the $p_T$-dependence of the results. The $v_2(p_T)$ shows a clear difference between initially isotropic and anisotropic cases, independent of the two choices for  the initial central temperature. The difference appears at $p_T$ values higher than the peak position, around $p_T \approx 2$ GeV.

\section{Conclusions}

In this paper we calculated the differential yields and elliptic flow of in-medium dileptons from a momentum-anisotropic QGP generated in Pb-Pb collision at $\sqrt{s}=2.76\ {\rm TeV}$ at LHC. We used 3+1d relativistic anisotropic hydrodynamics to model the non-equilibrium dynamics of the QGP  and convolved it with corresponding local rest frame dilepton emission rate from quarks and anti-quarks with ellipsoidally anisotropic momentum distributions. We presented the yield and flow results for intermediate mass dileptons with different settings of the aHydro parameters. The parameters taken for the background hydrodynamic evolution, e.g. shear viscosity-entropy density ratio  $\eta/s$ and initial temperature, were set based on their fitness in reproducing soft hadron spectra at the corresponding collision energy. 

The importance of corrections to the yield and flow due to the LRF momentum anisotropies was shown by comparing to the results of aHydro combined with isotropic LRF rates and also to the results of low dissipation/low anisotropy hydrodynamic evolution. We found that momentum-anisotropic corrections to dilepton emission rate have significant effects on the results and their interpretation.

Comparing the results for different values of $\eta/s$, we found that the intermediate invariant mass/transverse momentum QGP dilepton yield was always higher for smaller $\eta/s$ (corresponding to higher $T_0$). On the other hand, for dilepton elliptic flow, while for $M \lesssim 2\ {\rm GeV}$ and $p_T \lesssim 4\ {\rm GeV}$ the $v_2$ values for smaller $\eta/s$ were higher, the order was found to be reversed for $M \gtrsim 2\ {\rm GeV}$ and $p_T \gtrsim 4\ {\rm GeV}$ i.e. for harder dileptons, the $v_2$ was higher for QGP with larger $\eta/s$. 

The first order of momentum anisotropic correction to the distribution functions is the additive shear viscous term which, in general, reduces the elliptic flow compared to the ideal isotropic case \cite{Vujanovic:2013jpa, Heinz:2013th, Romatschke:2007mq, Dusling:2009df}. With the anisotropic parameterization \eqref{ellipsoidalf} used in aHydro, for lower $p_T$ we expect similar behavior i.e. the suppression of elliptic flow, but for higher pT particles produced at highly anisotropic earlier stages it is possible that the LRF momentum anisotropic distribution functions introduce the opposite effect and increase the elliptic flow.
 
We presented our predictions for different centrality classes, finding cases with negative values of $v_2$ for central collisions. However, lacking fluctuating initial conditions in the aHydro model, the interpretation of results for central collisions would be premature. 
We also investigated the effects of initial momentum anisotropy on  dilepton yield and flow. At large transverse momentum, the results of the dilepton yield and $v_2(p_T)$ were sensitive to the initial momentum anisotropy, while the results of $v_2(M)$ were not. 

Regarding the experimental feasibility of extracting the QGP-produced dileptons there are two additional sources that contribute in the intermediate mass region \mbox{$1 \lesssim M \lesssim 3$} :  (1) dileptons from open heavy flavor decays \cite{Rapp:2013nxa,Song:2018xca} and (2) jet conversion \cite{Turbide:2006mc, Fu:2014daa}.  Dileptons from open heavy flavor decays dominate in the intermediate mass regime, but using information about the distance of closest approach (DCA) of the production vertices one can eliminate these falsely triggered dileptons from the data.  Such approach will be used by the ALICE collaboration in Run 3 \cite{Vorobyev:2018}.  At RHIC energies it may be possible to perform a similar open heavy flavor subtraction using STAR's newly commissioned Heavy Flavor Tracker \cite{Contin:2016vgc}.  Turning to the second source, jet conversion, we note that if one integrates over all transverse momentum, the low transvserse momentum part of the yields dominate.  In this case, the dilepton yield from jet conversion is sub-leading in the intermediate mass regime compare to QGP-produced dileptons \cite{Fu:2014daa}.  Taken together with the ability to experimentally subtract the open charm decay background, this gives some hope to be able to cleanly extract the QGP-produced dileptons in the intermediate mass regime and, hence, information about the initial state of the QGP.

There are many directions one can take to further improve our analysis. The results in this paper were focused only on mid-rapidity region of dilepton production. In future studies, the effects of variation with rapidity can be investigated. The rapidity dependence of dilepton emission is expected to be more sensitive to initial momentum anisotropy \citep{Mauricio:2007vz, Martinez:2008di, Martinez:2008mc, Ryblewski:2015hea}.
In this paper, the contributions from hadronic sources of dileptons were not included. If we want to investigate low-mass dilepton production/flow, we must also consider the hadronic sources, in-medium $\rho$ meson spectral modification, and smooth transition of phases to lower temperatures. 
Another caveat is that we only analyzed a very limited set of cases when considering the effects of initial momentum anisotropy. The initial momentum anisotropy in azimuthal direction $({\rm with}\ \displaystyle \frac{\alpha_x}{\alpha_y}\neq 1)$ is an interesting issue which requires future work, however, we currently do not have strong constraints on the magnitude of these azimuthal LRF anisotropies. When calculating dilepton rates, we also neglected the masses of partons both in the cross section and in the distribution functions. In a more complete analysis, the effects of intrinsic and in-medium masses of partons need to be considered. Going beyond leading-order it is important to apply the same framework developed here to NLO dilepton production, dilepton production via jet conversion, and possible polarization observables \cite{Baym:2017qxy}. 
Finally, one must investigate the effects of fluctuating initial conditions in anisotropic hydrodynamics, temperature dependent $\eta/s$, and additional dissipative corrections to the ellipsoidally anisotropic momentum distributions. Similar studies also need to be performed for real photons produced in heavy-ion collisions. These are left for future work. 

\section*{Acknowledgement}

B. S. Kasmaei and M. Strickland were supported by
the U.S. Department of Energy, Office of Science, Office of Nuclear Physics under Award No. DE-SC0013470.

\bibliography{dilepton_paper}

\begin{thebibliography}{90}%
\makeatletter
\providecommand \@ifxundefined [1]{%
 \@ifx{#1\undefined}
}%
\providecommand \@ifnum [1]{%
 \ifnum #1\expandafter \@firstoftwo
 \else \expandafter \@secondoftwo
 \fi
}%
\providecommand \@ifx [1]{%
 \ifx #1\expandafter \@firstoftwo
 \else \expandafter \@secondoftwo
 \fi
}%
\providecommand \natexlab [1]{#1}%
\providecommand \enquote  [1]{``#1''}%
\providecommand \bibnamefont  [1]{#1}%
\providecommand \bibfnamefont [1]{#1}%
\providecommand \citenamefont [1]{#1}%
\providecommand \href@noop [0]{\@secondoftwo}%
\providecommand \href [0]{\begingroup \@sanitize@url \@href}%
\providecommand \@href[1]{\@@startlink{#1}\@@href}%
\providecommand \@@href[1]{\endgroup#1\@@endlink}%
\providecommand \@sanitize@url [0]{\catcode `\\12\catcode `\$12\catcode
  `\&12\catcode `\#12\catcode `\^12\catcode `\_12\catcode `\%12\relax}%
\providecommand \@@startlink[1]{}%
\providecommand \@@endlink[0]{}%
\providecommand \url  [0]{\begingroup\@sanitize@url \@url }%
\providecommand \@url [1]{\endgroup\@href {#1}{\urlprefix }}%
\providecommand \urlprefix  [0]{URL }%
\providecommand \Eprint [0]{\href }%
\providecommand \doibase [0]{http://dx.doi.org/}%
\providecommand \selectlanguage [0]{\@gobble}%
\providecommand \bibinfo  [0]{\@secondoftwo}%
\providecommand \bibfield  [0]{\@secondoftwo}%
\providecommand \translation [1]{[#1]}%
\providecommand \BibitemOpen [0]{}%
\providecommand \bibitemStop [0]{}%
\providecommand \bibitemNoStop [0]{.\EOS\space}%
\providecommand \EOS [0]{\spacefactor3000\relax}%
\providecommand \BibitemShut  [1]{\csname bibitem#1\endcsname}%
\let\auto@bib@innerbib\@empty
\bibitem [{\citenamefont {Braun-Munzinger}\ \emph {et~al.}(2016)\citenamefont
  {Braun-Munzinger}, \citenamefont {Koch}, \citenamefont {Sch{\"a}fer},\ and\
  \citenamefont {Stachel}}]{Braun-Munzinger:2015hba}%
  \BibitemOpen
  \bibfield  {author} {\bibinfo {author} {\bibfnamefont {P.}~\bibnamefont
  {Braun-Munzinger}}, \bibinfo {author} {\bibfnamefont {V.}~\bibnamefont
  {Koch}}, \bibinfo {author} {\bibfnamefont {T.}~\bibnamefont {Sch{\"a}fer}}, \
  and\ \bibinfo {author} {\bibfnamefont {J.}~\bibnamefont {Stachel}},\ }\href
  {\doibase 10.1016/j.physrep.2015.12.003} {\bibfield  {journal} {\bibinfo
  {journal} {Phys. Rept.}\ }\textbf {\bibinfo {volume} {621}},\ \bibinfo
  {pages} {76} (\bibinfo {year} {2016})},\ \Eprint
  {http://arxiv.org/abs/1510.00442} {arXiv:1510.00442 [nucl-th]} \BibitemShut
  {NoStop}%
\bibitem [{\citenamefont {Shuryak}(2017)}]{Shuryak:2014zxa}%
  \BibitemOpen
  \bibfield  {author} {\bibinfo {author} {\bibfnamefont {E.}~\bibnamefont
  {Shuryak}},\ }\href {\doibase 10.1103/RevModPhys.89.035001} {\bibfield
  {journal} {\bibinfo  {journal} {Rev. Mod. Phys.}\ }\textbf {\bibinfo {volume}
  {89}},\ \bibinfo {pages} {035001} (\bibinfo {year} {2017})},\ \Eprint
  {http://arxiv.org/abs/1412.8393} {arXiv:1412.8393 [hep-ph]} \BibitemShut
  {NoStop}%
\bibitem [{\citenamefont {Schukraft}(2017)}]{Schukraft:2017nbn}%
  \BibitemOpen
  \bibfield  {author} {\bibinfo {author} {\bibfnamefont {J.}~\bibnamefont
  {Schukraft}},\ }\bibfield  {booktitle} {\emph {\bibinfo {booktitle}
  {{Proceedings, 26th International Conference on Ultrarelativistic
  Nucleus-Nucleus Collisions (Quark Matter 2017): Chicago,Illinois, USA,
  February 6-11, 2017}}},\ }\href {\doibase 10.1016/j.nuclphysa.2017.05.036}
  {\bibfield  {journal} {\bibinfo  {journal} {Nucl. Phys.}\ }\textbf {\bibinfo
  {volume} {A967}},\ \bibinfo {pages} {1} (\bibinfo {year} {2017})},\ \Eprint
  {http://arxiv.org/abs/1705.02646} {arXiv:1705.02646 [hep-ex]} \BibitemShut
  {NoStop}%
\bibitem [{\citenamefont {Ploskon}(2018)}]{Ploskon:2018yiy}%
  \BibitemOpen
  \bibfield  {author} {\bibinfo {author} {\bibfnamefont {M.}~\bibnamefont
  {Ploskon}},\ }in\ \href@noop {} {\emph {\bibinfo {booktitle} {{14th
  International Workshop on Hadron Physics (Hadron Physics 2018) Florianopolis,
  Santa Catarina, Brazil, March 18-23, 2018}}}}\ (\bibinfo {year} {2018})\
  \Eprint {http://arxiv.org/abs/1808.01411} {arXiv:1808.01411 [hep-ex]}
  \BibitemShut {NoStop}%
\bibitem [{\citenamefont {Ollitrault}(2008)}]{Ollitrault:2008zz}%
  \BibitemOpen
  \bibfield  {author} {\bibinfo {author} {\bibfnamefont {J.-Y.}\ \bibnamefont
  {Ollitrault}},\ }\bibfield  {booktitle} {\emph {\bibinfo {booktitle}
  {{Advanced School on Quark-Gluon Plasma Mumbai, India, July 3-13, 2007}}},\
  }\href {\doibase 10.1088/0143-0807/29/2/010} {\bibfield  {journal} {\bibinfo
  {journal} {Eur. J. Phys.}\ }\textbf {\bibinfo {volume} {29}},\ \bibinfo
  {pages} {275} (\bibinfo {year} {2008})},\ \Eprint
  {http://arxiv.org/abs/0708.2433} {arXiv:0708.2433 [nucl-th]} \BibitemShut
  {NoStop}%
\bibitem [{\citenamefont {Gale}\ \emph {et~al.}(2013)\citenamefont {Gale},
  \citenamefont {Jeon},\ and\ \citenamefont {Schenke}}]{Gale:2013da}%
  \BibitemOpen
  \bibfield  {author} {\bibinfo {author} {\bibfnamefont {C.}~\bibnamefont
  {Gale}}, \bibinfo {author} {\bibfnamefont {S.}~\bibnamefont {Jeon}}, \ and\
  \bibinfo {author} {\bibfnamefont {B.}~\bibnamefont {Schenke}},\ }\href
  {\doibase 10.1142/S0217751X13400113} {\bibfield  {journal} {\bibinfo
  {journal} {Int. J. Mod. Phys.}\ }\textbf {\bibinfo {volume} {A28}},\ \bibinfo
  {pages} {1340011} (\bibinfo {year} {2013})},\ \Eprint
  {http://arxiv.org/abs/1301.5893} {arXiv:1301.5893 [nucl-th]} \BibitemShut
  {NoStop}%
\bibitem [{\citenamefont {Jeon}\ and\ \citenamefont
  {Heinz}(2015)}]{Jeon:2015dfa}%
  \BibitemOpen
  \bibfield  {author} {\bibinfo {author} {\bibfnamefont {S.}~\bibnamefont
  {Jeon}}\ and\ \bibinfo {author} {\bibfnamefont {U.}~\bibnamefont {Heinz}},\
  }\href {\doibase 10.1142/S0218301315300106} {\bibfield  {journal} {\bibinfo
  {journal} {Int. J. Mod. Phys.}\ }\textbf {\bibinfo {volume} {E24}},\ \bibinfo
  {pages} {1530010} (\bibinfo {year} {2015})},\ \Eprint
  {http://arxiv.org/abs/1503.03931} {arXiv:1503.03931 [hep-ph]} \BibitemShut
  {NoStop}%
\bibitem [{\citenamefont {Alqahtani}\ \emph {et~al.}(2018)\citenamefont
  {Alqahtani}, \citenamefont {Nopoush},\ and\ \citenamefont
  {Strickland}}]{Alqahtani:2017mhy}%
  \BibitemOpen
  \bibfield  {author} {\bibinfo {author} {\bibfnamefont {M.}~\bibnamefont
  {Alqahtani}}, \bibinfo {author} {\bibfnamefont {M.}~\bibnamefont {Nopoush}},
  \ and\ \bibinfo {author} {\bibfnamefont {M.}~\bibnamefont {Strickland}},\
  }\href {\doibase 10.1016/j.ppnp.2018.05.004} {\bibfield  {journal} {\bibinfo
  {journal} {Prog. Part. Nucl. Phys.}\ }\textbf {\bibinfo {volume} {101}},\
  \bibinfo {pages} {204} (\bibinfo {year} {2018})},\ \Eprint
  {http://arxiv.org/abs/1712.03282} {arXiv:1712.03282 [nucl-th]} \BibitemShut
  {NoStop}%
\bibitem [{\citenamefont {Yagi}\ \emph {et~al.}(2005)\citenamefont {Yagi},
  \citenamefont {Hatsuda},\ and\ \citenamefont {Miake}}]{Yagi-Hatsuda}%
  \BibitemOpen
  \bibfield  {author} {\bibinfo {author} {\bibfnamefont {K.}~\bibnamefont
  {Yagi}}, \bibinfo {author} {\bibfnamefont {T.}~\bibnamefont {Hatsuda}}, \
  and\ \bibinfo {author} {\bibfnamefont {Y.}~\bibnamefont {Miake}},\
  }\href@noop {} {\emph {\bibinfo {title} {Quark-gluon plasma: From big bang to
  little bang}}},\ Vol.~\bibinfo {volume} {23}\ (\bibinfo  {publisher}
  {Cambridge University Press},\ \bibinfo {year} {2005})\BibitemShut {NoStop}%
\bibitem [{\citenamefont {Shuryak}(1978)}]{Shuryak:1978ij}%
  \BibitemOpen
  \bibfield  {author} {\bibinfo {author} {\bibfnamefont {E.~V.}\ \bibnamefont
  {Shuryak}},\ }\href {\doibase 10.1016/0370-2693(78)90370-2} {\bibfield
  {journal} {\bibinfo  {journal} {Phys. Lett.}\ }\textbf {\bibinfo {volume}
  {78B}},\ \bibinfo {pages} {150} (\bibinfo {year} {1978})},\ \bibinfo {note}
  {[Yad. Fiz.28,796(1978)]}\BibitemShut {NoStop}%
\bibitem [{\citenamefont {Bass}\ \emph {et~al.}(1999)\citenamefont {Bass},
  \citenamefont {Gyulassy}, \citenamefont {Stoecker},\ and\ \citenamefont
  {Greiner}}]{Bass:1998vz}%
  \BibitemOpen
  \bibfield  {author} {\bibinfo {author} {\bibfnamefont {S.~A.}\ \bibnamefont
  {Bass}}, \bibinfo {author} {\bibfnamefont {M.}~\bibnamefont {Gyulassy}},
  \bibinfo {author} {\bibfnamefont {H.}~\bibnamefont {Stoecker}}, \ and\
  \bibinfo {author} {\bibfnamefont {W.}~\bibnamefont {Greiner}},\ }\href
  {\doibase 10.1088/0954-3899/25/3/013} {\bibfield  {journal} {\bibinfo
  {journal} {J. Phys.}\ }\textbf {\bibinfo {volume} {G25}},\ \bibinfo {pages}
  {R1} (\bibinfo {year} {1999})},\ \Eprint
  {http://arxiv.org/abs/hep-ph/9810281} {arXiv:hep-ph/9810281 [hep-ph]}
  \BibitemShut {NoStop}%
\bibitem [{\citenamefont {McLerran}\ and\ \citenamefont
  {Toimela}(1985)}]{McLerran:1984ay}%
  \BibitemOpen
  \bibfield  {author} {\bibinfo {author} {\bibfnamefont {L.~D.}\ \bibnamefont
  {McLerran}}\ and\ \bibinfo {author} {\bibfnamefont {T.}~\bibnamefont
  {Toimela}},\ }\href {\doibase 10.1103/PhysRevD.31.545} {\bibfield  {journal}
  {\bibinfo  {journal} {Phys. Rev.}\ }\textbf {\bibinfo {volume} {D31}},\
  \bibinfo {pages} {545} (\bibinfo {year} {1985})}\BibitemShut {NoStop}%
\bibitem [{\citenamefont {Gale}\ and\ \citenamefont
  {Kapusta}(1987)}]{Gale:1987ki}%
  \BibitemOpen
  \bibfield  {author} {\bibinfo {author} {\bibfnamefont {C.}~\bibnamefont
  {Gale}}\ and\ \bibinfo {author} {\bibfnamefont {J.~I.}\ \bibnamefont
  {Kapusta}},\ }\href {\doibase 10.1103/PhysRevC.35.2107} {\bibfield  {journal}
  {\bibinfo  {journal} {Phys. Rev.}\ }\textbf {\bibinfo {volume} {C35}},\
  \bibinfo {pages} {2107} (\bibinfo {year} {1987})}\BibitemShut {NoStop}%
\bibitem [{\citenamefont {Strickland}(1994)}]{STRICKLAND1994245}%
  \BibitemOpen
  \bibfield  {author} {\bibinfo {author} {\bibfnamefont {M.}~\bibnamefont
  {Strickland}},\ }\href {\doibase
  https://doi.org/10.1016/0370-2693(94)91045-6} {\bibfield  {journal} {\bibinfo
   {journal} {Physics Letters B}\ }\textbf {\bibinfo {volume} {331}},\ \bibinfo
  {pages} {245 } (\bibinfo {year} {1994})}\BibitemShut {NoStop}%
\bibitem [{\citenamefont {Gale}(2013)}]{Gale:2012xq}%
  \BibitemOpen
  \bibfield  {author} {\bibinfo {author} {\bibfnamefont {C.}~\bibnamefont
  {Gale}},\ }\bibfield  {booktitle} {\emph {\bibinfo {booktitle} {{Proceedings,
  5th International Conference on Hard and Electromagnetic Probes of
  High-Energy Nuclear Collisions (Hard Probes 2012): Cagliari, Italy, May
  27-June 1, 2012}}},\ }\href {\doibase 10.1016/j.nuclphysa.2012.12.034}
  {\bibfield  {journal} {\bibinfo  {journal} {Nucl. Phys.}\ }\textbf {\bibinfo
  {volume} {A910-911}},\ \bibinfo {pages} {147} (\bibinfo {year} {2013})},\
  \Eprint {http://arxiv.org/abs/1208.2289} {arXiv:1208.2289 [hep-ph]}
  \BibitemShut {NoStop}%
\bibitem [{\citenamefont {Linnyk}\ \emph {et~al.}(2013)\citenamefont {Linnyk},
  \citenamefont {Cassing}, \citenamefont {Manninen}, \citenamefont
  {Bratkovskaya}, \citenamefont {Gossiaux}, \citenamefont {Aichelin},
  \citenamefont {Song},\ and\ \citenamefont {Ko}}]{Linnyk:2012pu}%
  \BibitemOpen
  \bibfield  {author} {\bibinfo {author} {\bibfnamefont {O.}~\bibnamefont
  {Linnyk}}, \bibinfo {author} {\bibfnamefont {W.}~\bibnamefont {Cassing}},
  \bibinfo {author} {\bibfnamefont {J.}~\bibnamefont {Manninen}}, \bibinfo
  {author} {\bibfnamefont {E.~L.}\ \bibnamefont {Bratkovskaya}}, \bibinfo
  {author} {\bibfnamefont {P.~B.}\ \bibnamefont {Gossiaux}}, \bibinfo {author}
  {\bibfnamefont {J.}~\bibnamefont {Aichelin}}, \bibinfo {author}
  {\bibfnamefont {T.}~\bibnamefont {Song}}, \ and\ \bibinfo {author}
  {\bibfnamefont {C.~M.}\ \bibnamefont {Ko}},\ }\href {\doibase
  10.1103/PhysRevC.87.014905} {\bibfield  {journal} {\bibinfo  {journal} {Phys.
  Rev.}\ }\textbf {\bibinfo {volume} {C87}},\ \bibinfo {pages} {014905}
  (\bibinfo {year} {2013})},\ \Eprint {http://arxiv.org/abs/1208.1279}
  {arXiv:1208.1279 [nucl-th]} \BibitemShut {NoStop}%
\bibitem [{\citenamefont {Sakaguchi}(2015)}]{Sakaguchi:2014ewa}%
  \BibitemOpen
  \bibfield  {author} {\bibinfo {author} {\bibfnamefont {T.}~\bibnamefont
  {Sakaguchi}},\ }\href {\doibase 10.1007/s12043-015-0970-3} {\bibfield
  {journal} {\bibinfo  {journal} {Pramana}\ }\textbf {\bibinfo {volume} {84}},\
  \bibinfo {pages} {845} (\bibinfo {year} {2015})},\ \Eprint
  {http://arxiv.org/abs/1401.2481} {arXiv:1401.2481 [nucl-ex]} \BibitemShut
  {NoStop}%
\bibitem [{\citenamefont {Bratkovskaya}(2014)}]{Bratkovskaya:2014mva}%
  \BibitemOpen
  \bibfield  {author} {\bibinfo {author} {\bibfnamefont {E.~L.}\ \bibnamefont
  {Bratkovskaya}},\ }\bibfield  {booktitle} {\emph {\bibinfo {booktitle}
  {{Proceedings, 24th International Conference on Ultra-Relativistic
  Nucleus-Nucleus Collisions (Quark Matter 2014): Darmstadt, Germany, May
  19-24, 2014}}},\ }\href {\doibase 10.1016/j.nuclphysa.2014.09.088} {\bibfield
   {journal} {\bibinfo  {journal} {Nucl. Phys.}\ }\textbf {\bibinfo {volume}
  {A931}},\ \bibinfo {pages} {194} (\bibinfo {year} {2014})},\ \Eprint
  {http://arxiv.org/abs/1408.3674} {arXiv:1408.3674 [hep-ph]} \BibitemShut
  {NoStop}%
\bibitem [{\citenamefont {Bratkovskaya}\ \emph {et~al.}(2015)\citenamefont
  {Bratkovskaya}, \citenamefont {Linnyk},\ and\ \citenamefont
  {Cassing}}]{Bratkovskaya:2014toa}%
  \BibitemOpen
  \bibfield  {author} {\bibinfo {author} {\bibfnamefont {E.~L.}\ \bibnamefont
  {Bratkovskaya}}, \bibinfo {author} {\bibfnamefont {O.}~\bibnamefont
  {Linnyk}}, \ and\ \bibinfo {author} {\bibfnamefont {W.}~\bibnamefont
  {Cassing}},\ }\bibfield  {booktitle} {\emph {\bibinfo {booktitle}
  {{Proceedings, 3rd International Conference on New Frontiers in Physics
  (ICNFP 2014): Kolymbari, Crete, Greece, July 28-August 6, 2014}}},\ }\href
  {\doibase 10.1051/epjconf/20159501002} {\bibfield  {journal} {\bibinfo
  {journal} {EPJ Web Conf.}\ }\textbf {\bibinfo {volume} {95}},\ \bibinfo
  {pages} {01002} (\bibinfo {year} {2015})},\ \Eprint
  {http://arxiv.org/abs/1409.4190} {arXiv:1409.4190 [nucl-th]} \BibitemShut
  {NoStop}%
\bibitem [{\citenamefont {Endres}\ \emph {et~al.}(2016)\citenamefont {Endres},
  \citenamefont {van Hees},\ and\ \citenamefont {Bleicher}}]{Endres:2016tkg}%
  \BibitemOpen
  \bibfield  {author} {\bibinfo {author} {\bibfnamefont {S.}~\bibnamefont
  {Endres}}, \bibinfo {author} {\bibfnamefont {H.}~\bibnamefont {van Hees}}, \
  and\ \bibinfo {author} {\bibfnamefont {M.}~\bibnamefont {Bleicher}},\ }\href
  {\doibase 10.1103/PhysRevC.94.024912} {\bibfield  {journal} {\bibinfo
  {journal} {Phys. Rev.}\ }\textbf {\bibinfo {volume} {C94}},\ \bibinfo {pages}
  {024912} (\bibinfo {year} {2016})},\ \Eprint
  {http://arxiv.org/abs/1604.06415} {arXiv:1604.06415 [nucl-th]} \BibitemShut
  {NoStop}%
\bibitem [{\citenamefont {Shen}(2016)}]{Shen:2016odt}%
  \BibitemOpen
  \bibfield  {author} {\bibinfo {author} {\bibfnamefont {C.}~\bibnamefont
  {Shen}},\ }\bibfield  {booktitle} {\emph {\bibinfo {booktitle} {{Proceedings,
  25th International Conference on Ultra-Relativistic Nucleus-Nucleus
  Collisions (Quark Matter 2015): Kobe, Japan, September 27-October 3,
  2015}}},\ }\href {\doibase 10.1016/j.nuclphysa.2016.02.033} {\bibfield
  {journal} {\bibinfo  {journal} {Nucl. Phys.}\ }\textbf {\bibinfo {volume}
  {A956}},\ \bibinfo {pages} {184} (\bibinfo {year} {2016})},\ \Eprint
  {http://arxiv.org/abs/1601.02563} {arXiv:1601.02563 [nucl-th]} \BibitemShut
  {NoStop}%
\bibitem [{\citenamefont {Paquet}(2017)}]{Paquet:2017wji}%
  \BibitemOpen
  \bibfield  {author} {\bibinfo {author} {\bibfnamefont {J.-F.}\ \bibnamefont
  {Paquet}},\ }\bibfield  {booktitle} {\emph {\bibinfo {booktitle}
  {{Proceedings, 26th International Conference on Ultra-relativistic
  Nucleus-Nucleus Collisions (Quark Matter 2017): Chicago, Illinois, USA,
  February 5-11, 2017}}},\ }\href {\doibase 10.1016/j.nuclphysa.2017.06.003}
  {\bibfield  {journal} {\bibinfo  {journal} {Nucl. Phys.}\ }\textbf {\bibinfo
  {volume} {A967}},\ \bibinfo {pages} {184} (\bibinfo {year} {2017})},\ \Eprint
  {http://arxiv.org/abs/1704.07842} {arXiv:1704.07842 [nucl-th]} \BibitemShut
  {NoStop}%
\bibitem [{\citenamefont {Campbell}(2017)}]{Campbell:2017kbo}%
  \BibitemOpen
  \bibfield  {author} {\bibinfo {author} {\bibfnamefont {S.}~\bibnamefont
  {Campbell}},\ }\bibfield  {booktitle} {\emph {\bibinfo {booktitle}
  {{Proceedings, 26th International Conference on Ultra-relativistic
  Nucleus-Nucleus Collisions (Quark Matter 2017): Chicago, Illinois, USA,
  February 5-11, 2017}}},\ }\href {\doibase 10.1016/j.nuclphysa.2017.05.099}
  {\bibfield  {journal} {\bibinfo  {journal} {Nucl. Phys.}\ }\textbf {\bibinfo
  {volume} {A967}},\ \bibinfo {pages} {177} (\bibinfo {year} {2017})},\ \Eprint
  {http://arxiv.org/abs/1704.06307} {arXiv:1704.06307 [nucl-ex]} \BibitemShut
  {NoStop}%
\bibitem [{\citenamefont {Chanfray}\ \emph {et~al.}(1996)\citenamefont
  {Chanfray}, \citenamefont {Rapp},\ and\ \citenamefont
  {Wambach}}]{Rapp:1995zy}%
  \BibitemOpen
  \bibfield  {author} {\bibinfo {author} {\bibfnamefont {G.}~\bibnamefont
  {Chanfray}}, \bibinfo {author} {\bibfnamefont {R.}~\bibnamefont {Rapp}}, \
  and\ \bibinfo {author} {\bibfnamefont {J.}~\bibnamefont {Wambach}},\ }\href
  {\doibase 10.1103/PhysRevLett.76.368} {\bibfield  {journal} {\bibinfo
  {journal} {Phys. Rev. Lett.}\ }\textbf {\bibinfo {volume} {76}},\ \bibinfo
  {pages} {368} (\bibinfo {year} {1996})}\BibitemShut {NoStop}%
\bibitem [{\citenamefont {Bratkovskaya}\ and\ \citenamefont
  {Ko}(1999)}]{Bratkovskaya:1998pr}%
  \BibitemOpen
  \bibfield  {author} {\bibinfo {author} {\bibfnamefont {E.~L.}\ \bibnamefont
  {Bratkovskaya}}\ and\ \bibinfo {author} {\bibfnamefont {C.~M.}\ \bibnamefont
  {Ko}},\ }\href {\doibase 10.1016/S0370-2693(98)01500-7} {\bibfield  {journal}
  {\bibinfo  {journal} {Phys. Lett.}\ }\textbf {\bibinfo {volume} {B445}},\
  \bibinfo {pages} {265} (\bibinfo {year} {1999})},\ \Eprint
  {http://arxiv.org/abs/nucl-th/9809056} {arXiv:nucl-th/9809056 [nucl-th]}
  \BibitemShut {NoStop}%
\bibitem [{\citenamefont {Rapp}\ and\ \citenamefont
  {Wambach}(2000)}]{Rapp:1999ej}%
  \BibitemOpen
  \bibfield  {author} {\bibinfo {author} {\bibfnamefont {R.}~\bibnamefont
  {Rapp}}\ and\ \bibinfo {author} {\bibfnamefont {J.}~\bibnamefont {Wambach}},\
  }\href {\doibase 10.1007/0-306-47101-9_1} {\bibfield  {journal} {\bibinfo
  {journal} {Adv. Nucl. Phys.}\ }\textbf {\bibinfo {volume} {25}},\ \bibinfo
  {pages} {1} (\bibinfo {year} {2000})},\ \Eprint
  {http://arxiv.org/abs/hep-ph/9909229} {arXiv:hep-ph/9909229 [hep-ph]}
  \BibitemShut {NoStop}%
\bibitem [{\citenamefont {van Hees}\ and\ \citenamefont
  {Rapp}(2008)}]{vanHees:2007th}%
  \BibitemOpen
  \bibfield  {author} {\bibinfo {author} {\bibfnamefont {H.}~\bibnamefont {van
  Hees}}\ and\ \bibinfo {author} {\bibfnamefont {R.}~\bibnamefont {Rapp}},\
  }\href {\doibase 10.1016/j.nuclphysa.2008.03.009} {\bibfield  {journal}
  {\bibinfo  {journal} {Nucl. Phys.}\ }\textbf {\bibinfo {volume} {A806}},\
  \bibinfo {pages} {339} (\bibinfo {year} {2008})},\ \Eprint
  {http://arxiv.org/abs/0711.3444} {arXiv:0711.3444 [hep-ph]} \BibitemShut
  {NoStop}%
\bibitem [{\citenamefont {Rapp}\ and\ \citenamefont {van
  Hees}(2016)}]{Rapp:2016xzw}%
  \BibitemOpen
  \bibfield  {author} {\bibinfo {author} {\bibfnamefont {R.}~\bibnamefont
  {Rapp}}\ and\ \bibinfo {author} {\bibfnamefont {H.}~\bibnamefont {van
  Hees}},\ }\href {\doibase 10.1140/epja/i2016-16257-0} {\bibfield  {journal}
  {\bibinfo  {journal} {Eur. Phys. J.}\ }\textbf {\bibinfo {volume} {A52}},\
  \bibinfo {pages} {257} (\bibinfo {year} {2016})},\ \Eprint
  {http://arxiv.org/abs/1608.05279} {arXiv:1608.05279 [hep-ph]} \BibitemShut
  {NoStop}%
\bibitem [{\citenamefont {Weldon}(1990)}]{Weldon:1990iw}%
  \BibitemOpen
  \bibfield  {author} {\bibinfo {author} {\bibfnamefont {H.~A.}\ \bibnamefont
  {Weldon}},\ }\href {\doibase 10.1103/PhysRevD.42.2384} {\bibfield  {journal}
  {\bibinfo  {journal} {Phys. Rev.}\ }\textbf {\bibinfo {volume} {D42}},\
  \bibinfo {pages} {2384} (\bibinfo {year} {1990})}\BibitemShut {NoStop}%
\bibitem [{\citenamefont {Tuchin}(2013)}]{Tuchin:2013bda}%
  \BibitemOpen
  \bibfield  {author} {\bibinfo {author} {\bibfnamefont {K.}~\bibnamefont
  {Tuchin}},\ }\href {\doibase 10.1103/PhysRevC.88.024910} {\bibfield
  {journal} {\bibinfo  {journal} {Phys. Rev.}\ }\textbf {\bibinfo {volume}
  {C88}},\ \bibinfo {pages} {024910} (\bibinfo {year} {2013})},\ \Eprint
  {http://arxiv.org/abs/1305.0545} {arXiv:1305.0545 [nucl-th]} \BibitemShut
  {NoStop}%
\bibitem [{\citenamefont {Ghiglieri}\ and\ \citenamefont
  {Moore}(2014)}]{Ghiglieri:2014kma}%
  \BibitemOpen
  \bibfield  {author} {\bibinfo {author} {\bibfnamefont {J.}~\bibnamefont
  {Ghiglieri}}\ and\ \bibinfo {author} {\bibfnamefont {G.~D.}\ \bibnamefont
  {Moore}},\ }\href {\doibase 10.1007/JHEP12(2014)029} {\bibfield  {journal}
  {\bibinfo  {journal} {JHEP}\ }\textbf {\bibinfo {volume} {12}},\ \bibinfo
  {pages} {029} (\bibinfo {year} {2014})},\ \Eprint
  {http://arxiv.org/abs/1410.4203} {arXiv:1410.4203 [hep-ph]} \BibitemShut
  {NoStop}%
\bibitem [{\citenamefont {Basar}\ \emph {et~al.}(2014)\citenamefont {Basar},
  \citenamefont {Kharzeev},\ and\ \citenamefont {Shuryak}}]{Basar:2014swa}%
  \BibitemOpen
  \bibfield  {author} {\bibinfo {author} {\bibfnamefont {G.}~\bibnamefont
  {Basar}}, \bibinfo {author} {\bibfnamefont {D.~E.}\ \bibnamefont {Kharzeev}},
  \ and\ \bibinfo {author} {\bibfnamefont {E.~V.}\ \bibnamefont {Shuryak}},\
  }\href {\doibase 10.1103/PhysRevC.90.014905} {\bibfield  {journal} {\bibinfo
  {journal} {Phys. Rev.}\ }\textbf {\bibinfo {volume} {C90}},\ \bibinfo {pages}
  {014905} (\bibinfo {year} {2014})},\ \Eprint {http://arxiv.org/abs/1402.2286}
  {arXiv:1402.2286 [hep-ph]} \BibitemShut {NoStop}%
\bibitem [{\citenamefont {Burnier}\ and\ \citenamefont
  {Gastaldi}(2016)}]{Burnier:2015rka}%
  \BibitemOpen
  \bibfield  {author} {\bibinfo {author} {\bibfnamefont {Y.}~\bibnamefont
  {Burnier}}\ and\ \bibinfo {author} {\bibfnamefont {C.}~\bibnamefont
  {Gastaldi}},\ }\href {\doibase 10.1103/PhysRevC.93.044902} {\bibfield
  {journal} {\bibinfo  {journal} {Phys. Rev.}\ }\textbf {\bibinfo {volume}
  {C93}},\ \bibinfo {pages} {044902} (\bibinfo {year} {2016})},\ \Eprint
  {http://arxiv.org/abs/1508.06978} {arXiv:1508.06978 [nucl-th]} \BibitemShut
  {NoStop}%
\bibitem [{\citenamefont {Bandyopadhyay}\ \emph {et~al.}(2016)\citenamefont
  {Bandyopadhyay}, \citenamefont {Haque}, \citenamefont {Mustafa},\ and\
  \citenamefont {Strickland}}]{Bandyopadhyay:2015wua}%
  \BibitemOpen
  \bibfield  {author} {\bibinfo {author} {\bibfnamefont {A.}~\bibnamefont
  {Bandyopadhyay}}, \bibinfo {author} {\bibfnamefont {N.}~\bibnamefont
  {Haque}}, \bibinfo {author} {\bibfnamefont {M.~G.}\ \bibnamefont {Mustafa}},
  \ and\ \bibinfo {author} {\bibfnamefont {M.}~\bibnamefont {Strickland}},\
  }\href {\doibase 10.1103/PhysRevD.93.065004} {\bibfield  {journal} {\bibinfo
  {journal} {Phys. Rev.}\ }\textbf {\bibinfo {volume} {D93}},\ \bibinfo {pages}
  {065004} (\bibinfo {year} {2016})},\ \Eprint
  {http://arxiv.org/abs/1508.06249} {arXiv:1508.06249 [hep-ph]} \BibitemShut
  {NoStop}%
\bibitem [{\citenamefont {Hidaka}\ \emph {et~al.}(2015)\citenamefont {Hidaka},
  \citenamefont {Lin}, \citenamefont {Pisarski},\ and\ \citenamefont
  {Satow}}]{Hidaka:2015ima}%
  \BibitemOpen
  \bibfield  {author} {\bibinfo {author} {\bibfnamefont {Y.}~\bibnamefont
  {Hidaka}}, \bibinfo {author} {\bibfnamefont {S.}~\bibnamefont {Lin}},
  \bibinfo {author} {\bibfnamefont {R.~D.}\ \bibnamefont {Pisarski}}, \ and\
  \bibinfo {author} {\bibfnamefont {D.}~\bibnamefont {Satow}},\ }\href
  {\doibase 10.1007/JHEP10(2015)005} {\bibfield  {journal} {\bibinfo  {journal}
  {JHEP}\ }\textbf {\bibinfo {volume} {10}},\ \bibinfo {pages} {005} (\bibinfo
  {year} {2015})},\ \Eprint {http://arxiv.org/abs/1504.01770} {arXiv:1504.01770
  [hep-ph]} \BibitemShut {NoStop}%
\bibitem [{\citenamefont {Sadooghi}\ and\ \citenamefont
  {Taghinavaz}(2017)}]{Sadooghi:2016jyf}%
  \BibitemOpen
  \bibfield  {author} {\bibinfo {author} {\bibfnamefont {N.}~\bibnamefont
  {Sadooghi}}\ and\ \bibinfo {author} {\bibfnamefont {F.}~\bibnamefont
  {Taghinavaz}},\ }\href {\doibase 10.1016/j.aop.2016.11.008} {\bibfield
  {journal} {\bibinfo  {journal} {Annals Phys.}\ }\textbf {\bibinfo {volume}
  {376}},\ \bibinfo {pages} {218} (\bibinfo {year} {2017})},\ \Eprint
  {http://arxiv.org/abs/1601.04887} {arXiv:1601.04887 [hep-ph]} \BibitemShut
  {NoStop}%
\bibitem [{\citenamefont {Srivastava}\ \emph {et~al.}(2003)\citenamefont
  {Srivastava}, \citenamefont {Gale},\ and\ \citenamefont
  {Fries}}]{Srivastava:2002ic}%
  \BibitemOpen
  \bibfield  {author} {\bibinfo {author} {\bibfnamefont {D.~K.}\ \bibnamefont
  {Srivastava}}, \bibinfo {author} {\bibfnamefont {C.}~\bibnamefont {Gale}}, \
  and\ \bibinfo {author} {\bibfnamefont {R.~J.}\ \bibnamefont {Fries}},\ }\href
  {\doibase 10.1103/PhysRevC.67.034903} {\bibfield  {journal} {\bibinfo
  {journal} {Phys. Rev.}\ }\textbf {\bibinfo {volume} {C67}},\ \bibinfo {pages}
  {034903} (\bibinfo {year} {2003})},\ \Eprint
  {http://arxiv.org/abs/nucl-th/0209063} {arXiv:nucl-th/0209063 [nucl-th]}
  \BibitemShut {NoStop}%
\bibitem [{\citenamefont {Turbide}\ \emph
  {et~al.}(2006{\natexlab{a}})\citenamefont {Turbide}, \citenamefont {Gale},
  \citenamefont {Srivastava},\ and\ \citenamefont {Fries}}]{Turbide:2006mc}%
  \BibitemOpen
  \bibfield  {author} {\bibinfo {author} {\bibfnamefont {S.}~\bibnamefont
  {Turbide}}, \bibinfo {author} {\bibfnamefont {C.}~\bibnamefont {Gale}},
  \bibinfo {author} {\bibfnamefont {D.~K.}\ \bibnamefont {Srivastava}}, \ and\
  \bibinfo {author} {\bibfnamefont {R.~J.}\ \bibnamefont {Fries}},\ }\href
  {\doibase 10.1103/PhysRevC.74.014903} {\bibfield  {journal} {\bibinfo
  {journal} {Phys. Rev.}\ }\textbf {\bibinfo {volume} {C74}},\ \bibinfo {pages}
  {014903} (\bibinfo {year} {2006}{\natexlab{a}})},\ \Eprint
  {http://arxiv.org/abs/hep-ph/0601042} {arXiv:hep-ph/0601042 [hep-ph]}
  \BibitemShut {NoStop}%
\bibitem [{\citenamefont {Fu}\ and\ \citenamefont {Xi}(2015)}]{Fu:2014daa}%
  \BibitemOpen
  \bibfield  {author} {\bibinfo {author} {\bibfnamefont {Y.-P.}\ \bibnamefont
  {Fu}}\ and\ \bibinfo {author} {\bibfnamefont {Q.}~\bibnamefont {Xi}},\ }\href
  {\doibase 10.1103/PhysRevC.92.024914} {\bibfield  {journal} {\bibinfo
  {journal} {Phys. Rev.}\ }\textbf {\bibinfo {volume} {C92}},\ \bibinfo {pages}
  {024914} (\bibinfo {year} {2015})},\ \Eprint {http://arxiv.org/abs/1410.5044}
  {arXiv:1410.5044 [hep-ph]} \BibitemShut {NoStop}%
\bibitem [{\citenamefont {Mukherjee}\ \emph {et~al.}(2017)\citenamefont
  {Mukherjee}, \citenamefont {Mandal},\ and\ \citenamefont
  {Roy}}]{Mukherjee:2016sep}%
  \BibitemOpen
  \bibfield  {author} {\bibinfo {author} {\bibfnamefont {A.}~\bibnamefont
  {Mukherjee}}, \bibinfo {author} {\bibfnamefont {M.}~\bibnamefont {Mandal}}, \
  and\ \bibinfo {author} {\bibfnamefont {P.}~\bibnamefont {Roy}},\ }\href
  {\doibase 10.1140/epja/i2017-12265-x} {\bibfield  {journal} {\bibinfo
  {journal} {Eur. Phys. J.}\ }\textbf {\bibinfo {volume} {A53}},\ \bibinfo
  {pages} {81} (\bibinfo {year} {2017})},\ \Eprint
  {http://arxiv.org/abs/1604.08313} {arXiv:1604.08313 [hep-ph]} \BibitemShut
  {NoStop}%
\bibitem [{\citenamefont {Chatterjee}\ \emph {et~al.}(2007)\citenamefont
  {Chatterjee}, \citenamefont {Srivastava}, \citenamefont {Heinz},\ and\
  \citenamefont {Gale}}]{Chatterjee:2007xk}%
  \BibitemOpen
  \bibfield  {author} {\bibinfo {author} {\bibfnamefont {R.}~\bibnamefont
  {Chatterjee}}, \bibinfo {author} {\bibfnamefont {D.~K.}\ \bibnamefont
  {Srivastava}}, \bibinfo {author} {\bibfnamefont {U.~W.}\ \bibnamefont
  {Heinz}}, \ and\ \bibinfo {author} {\bibfnamefont {C.}~\bibnamefont {Gale}},\
  }\href {\doibase 10.1103/PhysRevC.75.054909} {\bibfield  {journal} {\bibinfo
  {journal} {Phys. Rev.}\ }\textbf {\bibinfo {volume} {C75}},\ \bibinfo {pages}
  {054909} (\bibinfo {year} {2007})},\ \Eprint
  {http://arxiv.org/abs/nucl-th/0702039} {arXiv:nucl-th/0702039 [nucl-th]}
  \BibitemShut {NoStop}%
\bibitem [{\citenamefont {Vujanovic}\ \emph {et~al.}(2014)\citenamefont
  {Vujanovic}, \citenamefont {Young}, \citenamefont {Schenke}, \citenamefont
  {Rapp}, \citenamefont {Jeon},\ and\ \citenamefont
  {Gale}}]{Vujanovic:2013jpa}%
  \BibitemOpen
  \bibfield  {author} {\bibinfo {author} {\bibfnamefont {G.}~\bibnamefont
  {Vujanovic}}, \bibinfo {author} {\bibfnamefont {C.}~\bibnamefont {Young}},
  \bibinfo {author} {\bibfnamefont {B.}~\bibnamefont {Schenke}}, \bibinfo
  {author} {\bibfnamefont {R.}~\bibnamefont {Rapp}}, \bibinfo {author}
  {\bibfnamefont {S.}~\bibnamefont {Jeon}}, \ and\ \bibinfo {author}
  {\bibfnamefont {C.}~\bibnamefont {Gale}},\ }\href {\doibase
  10.1103/PhysRevC.89.034904} {\bibfield  {journal} {\bibinfo  {journal} {Phys.
  Rev.}\ }\textbf {\bibinfo {volume} {C89}},\ \bibinfo {pages} {034904}
  (\bibinfo {year} {2014})},\ \Eprint {http://arxiv.org/abs/1312.0676}
  {arXiv:1312.0676 [nucl-th]} \BibitemShut {NoStop}%
\bibitem [{\citenamefont {Vujanovic}\ \emph {et~al.}(2016)\citenamefont
  {Vujanovic}, \citenamefont {Paquet}, \citenamefont {Denicol}, \citenamefont
  {Luzum}, \citenamefont {Jeon},\ and\ \citenamefont
  {Gale}}]{Vujanovic:2016anq}%
  \BibitemOpen
  \bibfield  {author} {\bibinfo {author} {\bibfnamefont {G.}~\bibnamefont
  {Vujanovic}}, \bibinfo {author} {\bibfnamefont {J.-F.}\ \bibnamefont
  {Paquet}}, \bibinfo {author} {\bibfnamefont {G.~S.}\ \bibnamefont {Denicol}},
  \bibinfo {author} {\bibfnamefont {M.}~\bibnamefont {Luzum}}, \bibinfo
  {author} {\bibfnamefont {S.}~\bibnamefont {Jeon}}, \ and\ \bibinfo {author}
  {\bibfnamefont {C.}~\bibnamefont {Gale}},\ }\href {\doibase
  10.1103/PhysRevC.94.014904} {\bibfield  {journal} {\bibinfo  {journal} {Phys.
  Rev.}\ }\textbf {\bibinfo {volume} {C94}},\ \bibinfo {pages} {014904}
  (\bibinfo {year} {2016})},\ \Eprint {http://arxiv.org/abs/1602.01455}
  {arXiv:1602.01455 [nucl-th]} \BibitemShut {NoStop}%
\bibitem [{\citenamefont {Bhattacharya}\ \emph {et~al.}(2016)\citenamefont
  {Bhattacharya}, \citenamefont {Ryblewski},\ and\ \citenamefont
  {Strickland}}]{Bhattacharya:2015ada}%
  \BibitemOpen
  \bibfield  {author} {\bibinfo {author} {\bibfnamefont {L.}~\bibnamefont
  {Bhattacharya}}, \bibinfo {author} {\bibfnamefont {R.}~\bibnamefont
  {Ryblewski}}, \ and\ \bibinfo {author} {\bibfnamefont {M.}~\bibnamefont
  {Strickland}},\ }\href {\doibase 10.1103/PhysRevD.93.065005} {\bibfield
  {journal} {\bibinfo  {journal} {Phys. Rev.}\ }\textbf {\bibinfo {volume}
  {D93}},\ \bibinfo {pages} {065005} (\bibinfo {year} {2016})},\ \Eprint
  {http://arxiv.org/abs/1507.06605} {arXiv:1507.06605 [hep-ph]} \BibitemShut
  {NoStop}%
\bibitem [{\citenamefont {Heinz}\ and\ \citenamefont
  {Snellings}(2013)}]{Heinz:2013th}%
  \BibitemOpen
  \bibfield  {author} {\bibinfo {author} {\bibfnamefont {U.}~\bibnamefont
  {Heinz}}\ and\ \bibinfo {author} {\bibfnamefont {R.}~\bibnamefont
  {Snellings}},\ }\href {\doibase 10.1146/annurev-nucl-102212-170540}
  {\bibfield  {journal} {\bibinfo  {journal} {Ann. Rev. Nucl. Part. Sci.}\
  }\textbf {\bibinfo {volume} {63}},\ \bibinfo {pages} {123} (\bibinfo {year}
  {2013})},\ \Eprint {http://arxiv.org/abs/1301.2826} {arXiv:1301.2826
  [nucl-th]} \BibitemShut {NoStop}%
\bibitem [{\citenamefont {Mr\'owczy\ifmmode~\acute{n}\else \'{n}\fi{}ski}\ and\
  \citenamefont {Thoma}(2000)}]{Mrowczynski:2000ed}%
  \BibitemOpen
  \bibfield  {author} {\bibinfo {author} {\bibfnamefont {S.}~\bibnamefont
  {Mr\'owczy\ifmmode~\acute{n}\else \'{n}\fi{}ski}}\ and\ \bibinfo {author}
  {\bibfnamefont {M.~H.}\ \bibnamefont {Thoma}},\ }\href {\doibase
  10.1103/PhysRevD.62.036011} {\bibfield  {journal} {\bibinfo  {journal} {Phys.
  Rev. D}\ }\textbf {\bibinfo {volume} {62}},\ \bibinfo {pages} {036011}
  (\bibinfo {year} {2000})}\BibitemShut {NoStop}%
\bibitem [{\citenamefont {Romatschke}\ and\ \citenamefont
  {Strickland}(2004)}]{Romatschke:2004jh}%
  \BibitemOpen
  \bibfield  {author} {\bibinfo {author} {\bibfnamefont {P.}~\bibnamefont
  {Romatschke}}\ and\ \bibinfo {author} {\bibfnamefont {M.}~\bibnamefont
  {Strickland}},\ }\href@noop {} {\bibfield  {journal} {\bibinfo  {journal}
  {Phys. Rev.}\ }\textbf {\bibinfo {volume} {D70}},\ \bibinfo {pages} {116006}
  (\bibinfo {year} {2004})},\ \Eprint {http://arxiv.org/abs/hep-ph/0406188}
  {hep-ph/0406188} \BibitemShut {NoStop}%
\bibitem [{\citenamefont {Martinez}\ and\ \citenamefont
  {Strickland}(2008{\natexlab{a}})}]{Mauricio:2007vz}%
  \BibitemOpen
  \bibfield  {author} {\bibinfo {author} {\bibfnamefont {M.}~\bibnamefont
  {Martinez}}\ and\ \bibinfo {author} {\bibfnamefont {M.}~\bibnamefont
  {Strickland}},\ }\href {\doibase 10.1103/PhysRevLett.100.102301} {\bibfield
  {journal} {\bibinfo  {journal} {Phys. Rev. Lett.}\ }\textbf {\bibinfo
  {volume} {100}},\ \bibinfo {pages} {102301} (\bibinfo {year}
  {2008}{\natexlab{a}})},\ \Eprint {http://arxiv.org/abs/0709.3576}
  {arXiv:0709.3576 [hep-ph]} \BibitemShut {NoStop}%
\bibitem [{\citenamefont {Martinez}\ and\ \citenamefont
  {Strickland}(2008{\natexlab{b}})}]{Martinez:2008di}%
  \BibitemOpen
  \bibfield  {author} {\bibinfo {author} {\bibfnamefont {M.}~\bibnamefont
  {Martinez}}\ and\ \bibinfo {author} {\bibfnamefont {M.}~\bibnamefont
  {Strickland}},\ }\href {\doibase 10.1103/PhysRevC.78.034917} {\bibfield
  {journal} {\bibinfo  {journal} {Phys. Rev.}\ }\textbf {\bibinfo {volume}
  {C78}},\ \bibinfo {pages} {034917} (\bibinfo {year} {2008}{\natexlab{b}})},\
  \Eprint {http://arxiv.org/abs/0805.4552} {arXiv:0805.4552 [hep-ph]}
  \BibitemShut {NoStop}%
\bibitem [{\citenamefont {Martinez}\ and\ \citenamefont
  {Strickland}(2009)}]{Martinez:2008mc}%
  \BibitemOpen
  \bibfield  {author} {\bibinfo {author} {\bibfnamefont {M.}~\bibnamefont
  {Martinez}}\ and\ \bibinfo {author} {\bibfnamefont {M.}~\bibnamefont
  {Strickland}},\ }\href {\doibase 10.1140/epjc/s10052-008-0851-8} {\bibfield
  {journal} {\bibinfo  {journal} {Eur. Phys. J.}\ }\textbf {\bibinfo {volume}
  {C61}},\ \bibinfo {pages} {905} (\bibinfo {year} {2009})},\ \Eprint
  {http://arxiv.org/abs/0808.3969} {arXiv:0808.3969 [hep-ph]} \BibitemShut
  {NoStop}%
\bibitem [{\citenamefont {Ryblewski}\ and\ \citenamefont
  {Strickland}(2015)}]{Ryblewski:2015hea}%
  \BibitemOpen
  \bibfield  {author} {\bibinfo {author} {\bibfnamefont {R.}~\bibnamefont
  {Ryblewski}}\ and\ \bibinfo {author} {\bibfnamefont {M.}~\bibnamefont
  {Strickland}},\ }\href {\doibase 10.1103/PhysRevD.92.025026} {\bibfield
  {journal} {\bibinfo  {journal} {Phys. Rev.}\ }\textbf {\bibinfo {volume}
  {D92}},\ \bibinfo {pages} {025026} (\bibinfo {year} {2015})},\ \Eprint
  {http://arxiv.org/abs/1501.03418} {arXiv:1501.03418 [nucl-th]} \BibitemShut
  {NoStop}%
\bibitem [{\citenamefont {Blaizot}\ and\ \citenamefont
  {Ollitrault}(1990)}]{Blaizot:1990zd}%
  \BibitemOpen
  \bibfield  {author} {\bibinfo {author} {\bibfnamefont {J.-P.}\ \bibnamefont
  {Blaizot}}\ and\ \bibinfo {author} {\bibfnamefont {J.-Y.}\ \bibnamefont
  {Ollitrault}},\ }\href {\doibase 10.1142/9789814503297_0008} {\bibfield
  {journal} {\bibinfo  {journal} {Adv. Ser. Direct. High Energy Phys.}\
  }\textbf {\bibinfo {volume} {6}},\ \bibinfo {pages} {393} (\bibinfo {year}
  {1990})}\BibitemShut {NoStop}%
\bibitem [{\citenamefont {Schenke}\ \emph {et~al.}(2011)\citenamefont
  {Schenke}, \citenamefont {Jeon},\ and\ \citenamefont
  {Gale}}]{Schenke:2010rr}%
  \BibitemOpen
  \bibfield  {author} {\bibinfo {author} {\bibfnamefont {B.}~\bibnamefont
  {Schenke}}, \bibinfo {author} {\bibfnamefont {S.}~\bibnamefont {Jeon}}, \
  and\ \bibinfo {author} {\bibfnamefont {C.}~\bibnamefont {Gale}},\ }\href
  {\doibase 10.1103/PhysRevLett.106.042301} {\bibfield  {journal} {\bibinfo
  {journal} {Phys. Rev. Lett.}\ }\textbf {\bibinfo {volume} {106}},\ \bibinfo
  {pages} {042301} (\bibinfo {year} {2011})},\ \Eprint
  {http://arxiv.org/abs/1009.3244} {arXiv:1009.3244 [hep-ph]} \BibitemShut
  {NoStop}%
\bibitem [{\citenamefont {Shen}\ \emph {et~al.}(2011)\citenamefont {Shen},
  \citenamefont {Heinz}, \citenamefont {Huovinen},\ and\ \citenamefont
  {Song}}]{Shen:2011eg}%
  \BibitemOpen
  \bibfield  {author} {\bibinfo {author} {\bibfnamefont {C.}~\bibnamefont
  {Shen}}, \bibinfo {author} {\bibfnamefont {U.}~\bibnamefont {Heinz}},
  \bibinfo {author} {\bibfnamefont {P.}~\bibnamefont {Huovinen}}, \ and\
  \bibinfo {author} {\bibfnamefont {H.}~\bibnamefont {Song}},\ }\href {\doibase
  10.1103/PhysRevC.84.044903} {\bibfield  {journal} {\bibinfo  {journal} {Phys.
  Rev.}\ }\textbf {\bibinfo {volume} {C84}},\ \bibinfo {pages} {044903}
  (\bibinfo {year} {2011})},\ \Eprint {http://arxiv.org/abs/1105.3226}
  {arXiv:1105.3226 [nucl-th]} \BibitemShut {NoStop}%
\bibitem [{\citenamefont {Alqahtani}\ \emph
  {et~al.}(2017{\natexlab{a}})\citenamefont {Alqahtani}, \citenamefont
  {Nopoush}, \citenamefont {Ryblewski},\ and\ \citenamefont
  {Strickland}}]{Alqahtani:2017jwl}%
  \BibitemOpen
  \bibfield  {author} {\bibinfo {author} {\bibfnamefont {M.}~\bibnamefont
  {Alqahtani}}, \bibinfo {author} {\bibfnamefont {M.}~\bibnamefont {Nopoush}},
  \bibinfo {author} {\bibfnamefont {R.}~\bibnamefont {Ryblewski}}, \ and\
  \bibinfo {author} {\bibfnamefont {M.}~\bibnamefont {Strickland}},\ }\href
  {\doibase 10.1103/PhysRevLett.119.042301} {\bibfield  {journal} {\bibinfo
  {journal} {Phys. Rev. Lett.}\ }\textbf {\bibinfo {volume} {119}},\ \bibinfo
  {pages} {042301} (\bibinfo {year} {2017}{\natexlab{a}})},\ \Eprint
  {http://arxiv.org/abs/1703.05808} {arXiv:1703.05808 [nucl-th]} \BibitemShut
  {NoStop}%
\bibitem [{\citenamefont {Romatschke}\ and\ \citenamefont
  {Romatschke}(2007)}]{Romatschke:2007mq}%
  \BibitemOpen
  \bibfield  {author} {\bibinfo {author} {\bibfnamefont {P.}~\bibnamefont
  {Romatschke}}\ and\ \bibinfo {author} {\bibfnamefont {U.}~\bibnamefont
  {Romatschke}},\ }\href {\doibase 10.1103/PhysRevLett.99.172301} {\bibfield
  {journal} {\bibinfo  {journal} {Phys. Rev. Lett.}\ }\textbf {\bibinfo
  {volume} {99}},\ \bibinfo {pages} {172301} (\bibinfo {year} {2007})},\
  \Eprint {http://arxiv.org/abs/0706.1522} {arXiv:0706.1522 [nucl-th]}
  \BibitemShut {NoStop}%
\bibitem [{\citenamefont {Dusling}\ and\ \citenamefont
  {Teaney}(2008)}]{Dusling:2007gi}%
  \BibitemOpen
  \bibfield  {author} {\bibinfo {author} {\bibfnamefont {K.}~\bibnamefont
  {Dusling}}\ and\ \bibinfo {author} {\bibfnamefont {D.}~\bibnamefont
  {Teaney}},\ }\href {\doibase 10.1103/PhysRevC.77.034905} {\bibfield
  {journal} {\bibinfo  {journal} {Phys. Rev.}\ }\textbf {\bibinfo {volume}
  {C77}},\ \bibinfo {pages} {034905} (\bibinfo {year} {2008})},\ \Eprint
  {http://arxiv.org/abs/0710.5932} {arXiv:0710.5932 [nucl-th]} \BibitemShut
  {NoStop}%
\bibitem [{\citenamefont {Luzum}\ and\ \citenamefont
  {Romatschke}(2008)}]{Luzum:2008cw}%
  \BibitemOpen
  \bibfield  {author} {\bibinfo {author} {\bibfnamefont {M.}~\bibnamefont
  {Luzum}}\ and\ \bibinfo {author} {\bibfnamefont {P.}~\bibnamefont
  {Romatschke}},\ }\href {\doibase 10.1103/PhysRevC.78.034915,
  10.1103/PhysRevC.79.039903} {\bibfield  {journal} {\bibinfo  {journal} {Phys.
  Rev.}\ }\textbf {\bibinfo {volume} {C78}},\ \bibinfo {pages} {034915}
  (\bibinfo {year} {2008})},\ \bibinfo {note} {[Erratum: Phys.
  Rev.C79,039903(2009)]},\ \Eprint {http://arxiv.org/abs/0804.4015}
  {arXiv:0804.4015 [nucl-th]} \BibitemShut {NoStop}%
\bibitem [{\citenamefont {Dusling}\ and\ \citenamefont
  {Lin}(2008)}]{Dusling:2008xj}%
  \BibitemOpen
  \bibfield  {author} {\bibinfo {author} {\bibfnamefont {K.}~\bibnamefont
  {Dusling}}\ and\ \bibinfo {author} {\bibfnamefont {S.}~\bibnamefont {Lin}},\
  }\href {\doibase 10.1016/j.nuclphysa.2008.06.007} {\bibfield  {journal}
  {\bibinfo  {journal} {Nucl. Phys.}\ }\textbf {\bibinfo {volume} {A809}},\
  \bibinfo {pages} {246} (\bibinfo {year} {2008})},\ \Eprint
  {http://arxiv.org/abs/0803.1262} {arXiv:0803.1262 [nucl-th]} \BibitemShut
  {NoStop}%
\bibitem [{\citenamefont {Romatschke}(2010)}]{Romatschke:2009im}%
  \BibitemOpen
  \bibfield  {author} {\bibinfo {author} {\bibfnamefont {P.}~\bibnamefont
  {Romatschke}},\ }\href {\doibase 10.1142/S0218301310014613} {\bibfield
  {journal} {\bibinfo  {journal} {Int. J. Mod. Phys.}\ }\textbf {\bibinfo
  {volume} {E19}},\ \bibinfo {pages} {1} (\bibinfo {year} {2010})},\ \Eprint
  {http://arxiv.org/abs/0902.3663} {arXiv:0902.3663 [hep-ph]} \BibitemShut
  {NoStop}%
\bibitem [{\citenamefont {Strickland}(2014)}]{Strickland:2014pga}%
  \BibitemOpen
  \bibfield  {author} {\bibinfo {author} {\bibfnamefont {M.}~\bibnamefont
  {Strickland}},\ }\bibfield  {booktitle} {\emph {\bibinfo {booktitle} {{54th
  Cracow School of Theoretical Physics: QCD meets experiment: Zakopane, Poland,
  June 12-20, 2014}}},\ }\href {\doibase 10.5506/APhysPolB.45.2355} {\bibfield
  {journal} {\bibinfo  {journal} {Acta Phys. Polon.}\ }\textbf {\bibinfo
  {volume} {B45}},\ \bibinfo {pages} {2355} (\bibinfo {year} {2014})},\ \Eprint
  {http://arxiv.org/abs/1410.5786} {arXiv:1410.5786 [nucl-th]} \BibitemShut
  {NoStop}%
\bibitem [{\citenamefont {Florkowski}\ and\ \citenamefont
  {Ryblewski}(2011)}]{Florkowski:2010cf}%
  \BibitemOpen
  \bibfield  {author} {\bibinfo {author} {\bibfnamefont {W.}~\bibnamefont
  {Florkowski}}\ and\ \bibinfo {author} {\bibfnamefont {R.}~\bibnamefont
  {Ryblewski}},\ }\href {\doibase 10.1103/PhysRevC.83.034907} {\bibfield
  {journal} {\bibinfo  {journal} {Phys. Rev.}\ }\textbf {\bibinfo {volume}
  {C83}},\ \bibinfo {pages} {034907} (\bibinfo {year} {2011})},\ \Eprint
  {http://arxiv.org/abs/1007.0130} {arXiv:1007.0130 [nucl-th]} \BibitemShut
  {NoStop}%
\bibitem [{\citenamefont {Ryblewski}\ and\ \citenamefont
  {Florkowski}(2011)}]{Ryblewski:2010bs}%
  \BibitemOpen
  \bibfield  {author} {\bibinfo {author} {\bibfnamefont {R.}~\bibnamefont
  {Ryblewski}}\ and\ \bibinfo {author} {\bibfnamefont {W.}~\bibnamefont
  {Florkowski}},\ }\href {\doibase 10.1088/0954-3899/38/1/015104} {\bibfield
  {journal} {\bibinfo  {journal} {J. Phys.}\ }\textbf {\bibinfo {volume}
  {G38}},\ \bibinfo {pages} {015104} (\bibinfo {year} {2011})},\ \Eprint
  {http://arxiv.org/abs/1007.4662} {arXiv:1007.4662 [nucl-th]} \BibitemShut
  {NoStop}%
\bibitem [{\citenamefont {Martinez}\ \emph {et~al.}(2012)\citenamefont
  {Martinez}, \citenamefont {Ryblewski},\ and\ \citenamefont
  {Strickland}}]{Martinez:2012tu}%
  \BibitemOpen
  \bibfield  {author} {\bibinfo {author} {\bibfnamefont {M.}~\bibnamefont
  {Martinez}}, \bibinfo {author} {\bibfnamefont {R.}~\bibnamefont {Ryblewski}},
  \ and\ \bibinfo {author} {\bibfnamefont {M.}~\bibnamefont {Strickland}},\
  }\href {\doibase 10.1103/PhysRevC.85.064913} {\bibfield  {journal} {\bibinfo
  {journal} {Phys. Rev.}\ }\textbf {\bibinfo {volume} {C85}},\ \bibinfo {pages}
  {064913} (\bibinfo {year} {2012})},\ \Eprint {http://arxiv.org/abs/1204.1473}
  {arXiv:1204.1473 [nucl-th]} \BibitemShut {NoStop}%
\bibitem [{\citenamefont {Nopoush}\ \emph {et~al.}(2014)\citenamefont
  {Nopoush}, \citenamefont {Ryblewski},\ and\ \citenamefont
  {Strickland}}]{Nopoush:2014pfa}%
  \BibitemOpen
  \bibfield  {author} {\bibinfo {author} {\bibfnamefont {M.}~\bibnamefont
  {Nopoush}}, \bibinfo {author} {\bibfnamefont {R.}~\bibnamefont {Ryblewski}},
  \ and\ \bibinfo {author} {\bibfnamefont {M.}~\bibnamefont {Strickland}},\
  }\href {\doibase 10.1103/PhysRevC.90.014908} {\bibfield  {journal} {\bibinfo
  {journal} {Phys. Rev.}\ }\textbf {\bibinfo {volume} {C90}},\ \bibinfo {pages}
  {014908} (\bibinfo {year} {2014})},\ \Eprint {http://arxiv.org/abs/1405.1355}
  {arXiv:1405.1355 [hep-ph]} \BibitemShut {NoStop}%
\bibitem [{\citenamefont {Florkowski}\ \emph {et~al.}(2016)\citenamefont
  {Florkowski}, \citenamefont {Ryblewski}, \citenamefont {Strickland},\ and\
  \citenamefont {Tinti}}]{Florkowski:2016kjj}%
  \BibitemOpen
  \bibfield  {author} {\bibinfo {author} {\bibfnamefont {W.}~\bibnamefont
  {Florkowski}}, \bibinfo {author} {\bibfnamefont {R.}~\bibnamefont
  {Ryblewski}}, \bibinfo {author} {\bibfnamefont {M.}~\bibnamefont
  {Strickland}}, \ and\ \bibinfo {author} {\bibfnamefont {L.}~\bibnamefont
  {Tinti}},\ }\href {\doibase 10.1103/PhysRevC.94.064903} {\bibfield  {journal}
  {\bibinfo  {journal} {Phys. Rev.}\ }\textbf {\bibinfo {volume} {C94}},\
  \bibinfo {pages} {064903} (\bibinfo {year} {2016})},\ \Eprint
  {http://arxiv.org/abs/1609.06293} {arXiv:1609.06293 [nucl-th]} \BibitemShut
  {NoStop}%
\bibitem [{\citenamefont {Gale}\ \emph {et~al.}(2015)\citenamefont {Gale},
  \citenamefont {Hidaka}, \citenamefont {Jeon}, \citenamefont {Lin},
  \citenamefont {Paquet}, \citenamefont {Pisarski}, \citenamefont {Satow},
  \citenamefont {Skokov},\ and\ \citenamefont {Vujanovic}}]{Gale:2014dfa}%
  \BibitemOpen
  \bibfield  {author} {\bibinfo {author} {\bibfnamefont {C.}~\bibnamefont
  {Gale}}, \bibinfo {author} {\bibfnamefont {Y.}~\bibnamefont {Hidaka}},
  \bibinfo {author} {\bibfnamefont {S.}~\bibnamefont {Jeon}}, \bibinfo {author}
  {\bibfnamefont {S.}~\bibnamefont {Lin}}, \bibinfo {author} {\bibfnamefont
  {J.-F.}\ \bibnamefont {Paquet}}, \bibinfo {author} {\bibfnamefont {R.~D.}\
  \bibnamefont {Pisarski}}, \bibinfo {author} {\bibfnamefont {D.}~\bibnamefont
  {Satow}}, \bibinfo {author} {\bibfnamefont {V.~V.}\ \bibnamefont {Skokov}}, \
  and\ \bibinfo {author} {\bibfnamefont {G.}~\bibnamefont {Vujanovic}},\ }\href
  {\doibase 10.1103/PhysRevLett.114.072301} {\bibfield  {journal} {\bibinfo
  {journal} {Phys. Rev. Lett.}\ }\textbf {\bibinfo {volume} {114}},\ \bibinfo
  {pages} {072301} (\bibinfo {year} {2015})},\ \Eprint
  {http://arxiv.org/abs/1409.4778} {arXiv:1409.4778 [hep-ph]} \BibitemShut
  {NoStop}%
\bibitem [{\citenamefont {Kajantie}\ \emph {et~al.}(1986)\citenamefont
  {Kajantie}, \citenamefont {Kapusta}, \citenamefont {McLerran},\ and\
  \citenamefont {Mekjian}}]{Kajantie:1986dh}%
  \BibitemOpen
  \bibfield  {author} {\bibinfo {author} {\bibfnamefont {K.}~\bibnamefont
  {Kajantie}}, \bibinfo {author} {\bibfnamefont {J.~I.}\ \bibnamefont
  {Kapusta}}, \bibinfo {author} {\bibfnamefont {L.~D.}\ \bibnamefont
  {McLerran}}, \ and\ \bibinfo {author} {\bibfnamefont {A.}~\bibnamefont
  {Mekjian}},\ }\href {\doibase 10.1103/PhysRevD.34.2746} {\bibfield  {journal}
  {\bibinfo  {journal} {Phys. Rev.}\ }\textbf {\bibinfo {volume} {D34}},\
  \bibinfo {pages} {2746} (\bibinfo {year} {1986})}\BibitemShut {NoStop}%
\bibitem [{\citenamefont {Ruuskanen}(1990)}]{Ruuskanen:1989tp}%
  \BibitemOpen
  \bibfield  {author} {\bibinfo {author} {\bibfnamefont {P.~V.}\ \bibnamefont
  {Ruuskanen}},\ }\href {\doibase 10.1142/9789814503297_0010} {\bibfield
  {journal} {\bibinfo  {journal} {Adv. Ser. Direct. High Energy Phys.}\
  }\textbf {\bibinfo {volume} {6}},\ \bibinfo {pages} {519} (\bibinfo {year}
  {1990})}\BibitemShut {NoStop}%
\bibitem [{\citenamefont {Cleymans}\ \emph {et~al.}(1987)\citenamefont
  {Cleymans}, \citenamefont {Fingberg},\ and\ \citenamefont
  {Redlich}}]{Cleymans:1986na}%
  \BibitemOpen
  \bibfield  {author} {\bibinfo {author} {\bibfnamefont {J.}~\bibnamefont
  {Cleymans}}, \bibinfo {author} {\bibfnamefont {J.}~\bibnamefont {Fingberg}},
  \ and\ \bibinfo {author} {\bibfnamefont {K.}~\bibnamefont {Redlich}},\ }\href
  {\doibase 10.1103/PhysRevD.35.2153} {\bibfield  {journal} {\bibinfo
  {journal} {Phys. Rev.}\ }\textbf {\bibinfo {volume} {D35}},\ \bibinfo {pages}
  {2153} (\bibinfo {year} {1987})}\BibitemShut {NoStop}%
\bibitem [{\citenamefont {Gorenstein}\ and\ \citenamefont
  {Mogilevsky}(1989)}]{Gorenstein:1989ks}%
  \BibitemOpen
  \bibfield  {author} {\bibinfo {author} {\bibfnamefont {M.~I.}\ \bibnamefont
  {Gorenstein}}\ and\ \bibinfo {author} {\bibfnamefont {O.~A.}\ \bibnamefont
  {Mogilevsky}},\ }\href {\doibase 10.1016/0370-2693(89)90536-4} {\bibfield
  {journal} {\bibinfo  {journal} {Phys. Lett.}\ }\textbf {\bibinfo {volume}
  {B228}},\ \bibinfo {pages} {121} (\bibinfo {year} {1989})}\BibitemShut
  {NoStop}%
\bibitem [{\citenamefont {Dumitru}\ \emph {et~al.}(1993)\citenamefont
  {Dumitru}, \citenamefont {Rischke}, \citenamefont {Sch\"onfeld},
  \citenamefont {Winckelmann}, \citenamefont {St\"ocker},\ and\ \citenamefont
  {Greiner}}]{Dumitru:1993vz}%
  \BibitemOpen
  \bibfield  {author} {\bibinfo {author} {\bibfnamefont {A.}~\bibnamefont
  {Dumitru}}, \bibinfo {author} {\bibfnamefont {D.~H.}\ \bibnamefont
  {Rischke}}, \bibinfo {author} {\bibfnamefont {T.}~\bibnamefont
  {Sch\"onfeld}}, \bibinfo {author} {\bibfnamefont {L.}~\bibnamefont
  {Winckelmann}}, \bibinfo {author} {\bibfnamefont {H.}~\bibnamefont
  {St\"ocker}}, \ and\ \bibinfo {author} {\bibfnamefont {W.}~\bibnamefont
  {Greiner}},\ }\href {\doibase 10.1103/PhysRevLett.70.2860} {\bibfield
  {journal} {\bibinfo  {journal} {Phys. Rev. Lett.}\ }\textbf {\bibinfo
  {volume} {70}},\ \bibinfo {pages} {2860} (\bibinfo {year}
  {1993})}\BibitemShut {NoStop}%
\bibitem [{\citenamefont {Alqahtani}\ \emph
  {et~al.}(2017{\natexlab{b}})\citenamefont {Alqahtani}, \citenamefont
  {Nopoush}, \citenamefont {Ryblewski},\ and\ \citenamefont
  {Strickland}}]{Alqahtani:2017tnq}%
  \BibitemOpen
  \bibfield  {author} {\bibinfo {author} {\bibfnamefont {M.}~\bibnamefont
  {Alqahtani}}, \bibinfo {author} {\bibfnamefont {M.}~\bibnamefont {Nopoush}},
  \bibinfo {author} {\bibfnamefont {R.}~\bibnamefont {Ryblewski}}, \ and\
  \bibinfo {author} {\bibfnamefont {M.}~\bibnamefont {Strickland}},\ }\href
  {\doibase 10.1103/PhysRevC.96.044910} {\bibfield  {journal} {\bibinfo
  {journal} {Phys. Rev.}\ }\textbf {\bibinfo {volume} {C96}},\ \bibinfo {pages}
  {044910} (\bibinfo {year} {2017}{\natexlab{b}})},\ \Eprint
  {http://arxiv.org/abs/1705.10191} {arXiv:1705.10191 [nucl-th]} \BibitemShut
  {NoStop}%
\bibitem [{\citenamefont {Miller}\ \emph {et~al.}(2007)\citenamefont {Miller},
  \citenamefont {Reygers}, \citenamefont {Sanders},\ and\ \citenamefont
  {Steinberg}}]{Miller:2007ri}%
  \BibitemOpen
  \bibfield  {author} {\bibinfo {author} {\bibfnamefont {M.~L.}\ \bibnamefont
  {Miller}}, \bibinfo {author} {\bibfnamefont {K.}~\bibnamefont {Reygers}},
  \bibinfo {author} {\bibfnamefont {S.~J.}\ \bibnamefont {Sanders}}, \ and\
  \bibinfo {author} {\bibfnamefont {P.}~\bibnamefont {Steinberg}},\ }\href
  {\doibase 10.1146/annurev.nucl.57.090506.123020} {\bibfield  {journal}
  {\bibinfo  {journal} {Ann. Rev. Nucl. Part. Sci.}\ }\textbf {\bibinfo
  {volume} {57}},\ \bibinfo {pages} {205} (\bibinfo {year} {2007})},\ \Eprint
  {http://arxiv.org/abs/nucl-ex/0701025} {arXiv:nucl-ex/0701025 [nucl-ex]}
  \BibitemShut {NoStop}%
\bibitem [{\citenamefont {Kovtun}\ \emph {et~al.}(2005)\citenamefont {Kovtun},
  \citenamefont {Son},\ and\ \citenamefont {Starinets}}]{Kovtun:2004de}%
  \BibitemOpen
  \bibfield  {author} {\bibinfo {author} {\bibfnamefont {P.~K.}\ \bibnamefont
  {Kovtun}}, \bibinfo {author} {\bibfnamefont {D.~T.}\ \bibnamefont {Son}}, \
  and\ \bibinfo {author} {\bibfnamefont {A.~O.}\ \bibnamefont {Starinets}},\
  }\href {\doibase 10.1103/PhysRevLett.94.111601} {\bibfield  {journal}
  {\bibinfo  {journal} {Phys. Rev. Lett.}\ }\textbf {\bibinfo {volume} {94}},\
  \bibinfo {pages} {111601} (\bibinfo {year} {2005})}\BibitemShut {NoStop}%
\bibitem [{\citenamefont {Borsanyi}\ \emph {et~al.}(2010)\citenamefont
  {Borsanyi}, \citenamefont {Endrodi}, \citenamefont {Fodor}, \citenamefont
  {Jakovac}, \citenamefont {Katz}, \citenamefont {Krieg}, \citenamefont
  {Ratti},\ and\ \citenamefont {Szabo}}]{Borsanyi:2010cj}%
  \BibitemOpen
  \bibfield  {author} {\bibinfo {author} {\bibfnamefont {S.}~\bibnamefont
  {Borsanyi}}, \bibinfo {author} {\bibfnamefont {G.}~\bibnamefont {Endrodi}},
  \bibinfo {author} {\bibfnamefont {Z.}~\bibnamefont {Fodor}}, \bibinfo
  {author} {\bibfnamefont {A.}~\bibnamefont {Jakovac}}, \bibinfo {author}
  {\bibfnamefont {S.~D.}\ \bibnamefont {Katz}}, \bibinfo {author}
  {\bibfnamefont {S.}~\bibnamefont {Krieg}}, \bibinfo {author} {\bibfnamefont
  {C.}~\bibnamefont {Ratti}}, \ and\ \bibinfo {author} {\bibfnamefont {K.~K.}\
  \bibnamefont {Szabo}},\ }\href {\doibase 10.1007/JHEP11(2010)077} {\bibfield
  {journal} {\bibinfo  {journal} {JHEP}\ }\textbf {\bibinfo {volume} {11}},\
  \bibinfo {pages} {077} (\bibinfo {year} {2010})},\ \Eprint
  {http://arxiv.org/abs/1007.2580} {arXiv:1007.2580 [hep-lat]} \BibitemShut
  {NoStop}%
\bibitem [{\citenamefont {Chojnacki}\ \emph {et~al.}(2012)\citenamefont
  {Chojnacki}, \citenamefont {Kisiel}, \citenamefont {Florkowski},\ and\
  \citenamefont {Broniowski}}]{Chojnacki:2011hb}%
  \BibitemOpen
  \bibfield  {author} {\bibinfo {author} {\bibfnamefont {M.}~\bibnamefont
  {Chojnacki}}, \bibinfo {author} {\bibfnamefont {A.}~\bibnamefont {Kisiel}},
  \bibinfo {author} {\bibfnamefont {W.}~\bibnamefont {Florkowski}}, \ and\
  \bibinfo {author} {\bibfnamefont {W.}~\bibnamefont {Broniowski}},\ }\href
  {\doibase 10.1016/j.cpc.2011.11.018} {\bibfield  {journal} {\bibinfo
  {journal} {Comput. Phys. Commun.}\ }\textbf {\bibinfo {volume} {183}},\
  \bibinfo {pages} {746} (\bibinfo {year} {2012})},\ \Eprint
  {http://arxiv.org/abs/1102.0273} {arXiv:1102.0273 [nucl-th]} \BibitemShut
  {NoStop}%
\bibitem [{\citenamefont {Danielewicz}\ \emph {et~al.}(1998)\citenamefont
  {Danielewicz}, \citenamefont {Lacey}, \citenamefont {Gossiaux}, \citenamefont
  {Pinkenburg}, \citenamefont {Chung}, \citenamefont {Alexander},\ and\
  \citenamefont {McGrath}}]{Danielewicz:1998vz}%
  \BibitemOpen
  \bibfield  {author} {\bibinfo {author} {\bibfnamefont {P.}~\bibnamefont
  {Danielewicz}}, \bibinfo {author} {\bibfnamefont {R.~A.}\ \bibnamefont
  {Lacey}}, \bibinfo {author} {\bibfnamefont {P.~B.}\ \bibnamefont {Gossiaux}},
  \bibinfo {author} {\bibfnamefont {C.}~\bibnamefont {Pinkenburg}}, \bibinfo
  {author} {\bibfnamefont {P.}~\bibnamefont {Chung}}, \bibinfo {author}
  {\bibfnamefont {J.~M.}\ \bibnamefont {Alexander}}, \ and\ \bibinfo {author}
  {\bibfnamefont {R.~L.}\ \bibnamefont {McGrath}},\ }\href {\doibase
  10.1103/PhysRevLett.81.2438} {\bibfield  {journal} {\bibinfo  {journal}
  {Phys. Rev. Lett.}\ }\textbf {\bibinfo {volume} {81}},\ \bibinfo {pages}
  {2438} (\bibinfo {year} {1998})},\ \Eprint
  {http://arxiv.org/abs/nucl-th/9803047} {arXiv:nucl-th/9803047 [nucl-th]}
  \BibitemShut {NoStop}%
\bibitem [{\citenamefont {Pinkenburg}\ \emph {et~al.}(1999)\citenamefont
  {Pinkenburg} \emph {et~al.}}]{Pinkenburg:1999ya}%
  \BibitemOpen
  \bibfield  {author} {\bibinfo {author} {\bibfnamefont {C.}~\bibnamefont
  {Pinkenburg}} \emph {et~al.} (\bibinfo {collaboration} {E895}),\ }\href
  {\doibase 10.1103/PhysRevLett.83.1295} {\bibfield  {journal} {\bibinfo
  {journal} {Phys. Rev. Lett.}\ }\textbf {\bibinfo {volume} {83}},\ \bibinfo
  {pages} {1295} (\bibinfo {year} {1999})},\ \Eprint
  {http://arxiv.org/abs/nucl-ex/9903010} {arXiv:nucl-ex/9903010 [nucl-ex]}
  \BibitemShut {NoStop}%
\bibitem [{\citenamefont {Turbide}\ \emph
  {et~al.}(2006{\natexlab{b}})\citenamefont {Turbide}, \citenamefont {Gale},\
  and\ \citenamefont {Fries}}]{Turbide:2005bz}%
  \BibitemOpen
  \bibfield  {author} {\bibinfo {author} {\bibfnamefont {S.}~\bibnamefont
  {Turbide}}, \bibinfo {author} {\bibfnamefont {C.}~\bibnamefont {Gale}}, \
  and\ \bibinfo {author} {\bibfnamefont {R.~J.}\ \bibnamefont {Fries}},\ }\href
  {\doibase 10.1103/PhysRevLett.96.032303} {\bibfield  {journal} {\bibinfo
  {journal} {Phys. Rev. Lett.}\ }\textbf {\bibinfo {volume} {96}},\ \bibinfo
  {pages} {032303} (\bibinfo {year} {2006}{\natexlab{b}})},\ \Eprint
  {http://arxiv.org/abs/hep-ph/0508201} {arXiv:hep-ph/0508201 [hep-ph]}
  \BibitemShut {NoStop}%
\bibitem [{\citenamefont {Krieg}\ and\ \citenamefont
  {Bleicher}(2009)}]{Krieg:2007bc}%
  \BibitemOpen
  \bibfield  {author} {\bibinfo {author} {\bibfnamefont {D.}~\bibnamefont
  {Krieg}}\ and\ \bibinfo {author} {\bibfnamefont {M.}~\bibnamefont
  {Bleicher}},\ }\href {\doibase 10.1140/epja/i2008-10700-9} {\bibfield
  {journal} {\bibinfo  {journal} {Eur. Phys. J.}\ }\textbf {\bibinfo {volume}
  {A39}},\ \bibinfo {pages} {1} (\bibinfo {year} {2009})},\ \Eprint
  {http://arxiv.org/abs/0806.0736} {arXiv:0806.0736 [nucl-th]} \BibitemShut
  {NoStop}%
\bibitem [{\citenamefont {Florkowski}\ \emph {et~al.}(2018)\citenamefont
  {Florkowski}, \citenamefont {Heller},\ and\ \citenamefont
  {Spalinski}}]{Florkowski:2017olj}%
  \BibitemOpen
  \bibfield  {author} {\bibinfo {author} {\bibfnamefont {W.}~\bibnamefont
  {Florkowski}}, \bibinfo {author} {\bibfnamefont {M.~P.}\ \bibnamefont
  {Heller}}, \ and\ \bibinfo {author} {\bibfnamefont {M.}~\bibnamefont
  {Spalinski}},\ }\href {\doibase 10.1088/1361-6633/aaa091} {\bibfield
  {journal} {\bibinfo  {journal} {Rept. Prog. Phys.}\ }\textbf {\bibinfo
  {volume} {81}},\ \bibinfo {pages} {046001} (\bibinfo {year} {2018})},\
  \Eprint {http://arxiv.org/abs/1707.02282} {arXiv:1707.02282 [hep-ph]}
  \BibitemShut {NoStop}%
\bibitem [{\citenamefont {Strickland}\ \emph {et~al.}(2018)\citenamefont
  {Strickland}, \citenamefont {Noronha},\ and\ \citenamefont
  {Denicol}}]{Strickland:2017kux}%
  \BibitemOpen
  \bibfield  {author} {\bibinfo {author} {\bibfnamefont {M.}~\bibnamefont
  {Strickland}}, \bibinfo {author} {\bibfnamefont {J.}~\bibnamefont {Noronha}},
  \ and\ \bibinfo {author} {\bibfnamefont {G.~S.}\ \bibnamefont {Denicol}},\
  }\href {\doibase 10.1103/PhysRevD.97.036020} {\bibfield  {journal} {\bibinfo
  {journal} {Phys. Rev.}\ }\textbf {\bibinfo {volume} {D97}},\ \bibinfo {pages}
  {036020} (\bibinfo {year} {2018})},\ \Eprint
  {http://arxiv.org/abs/1709.06644} {arXiv:1709.06644 [nucl-th]} \BibitemShut
  {NoStop}%
\bibitem [{\citenamefont {Abelev}\ \emph {et~al.}(2013)\citenamefont {Abelev}
  \emph {et~al.}}]{Abelev:2013vea}%
  \BibitemOpen
  \bibfield  {author} {\bibinfo {author} {\bibfnamefont {B.}~\bibnamefont
  {Abelev}} \emph {et~al.} (\bibinfo {collaboration} {ALICE}),\ }\href
  {\doibase 10.1103/PhysRevC.88.044910} {\bibfield  {journal} {\bibinfo
  {journal} {Phys. Rev.}\ }\textbf {\bibinfo {volume} {C88}},\ \bibinfo {pages}
  {044910} (\bibinfo {year} {2013})},\ \Eprint {http://arxiv.org/abs/1303.0737}
  {arXiv:1303.0737 [hep-ex]} \BibitemShut {NoStop}%
\bibitem [{\citenamefont {Dusling}\ \emph {et~al.}(2010)\citenamefont
  {Dusling}, \citenamefont {Moore},\ and\ \citenamefont
  {Teaney}}]{Dusling:2009df}%
  \BibitemOpen
  \bibfield  {author} {\bibinfo {author} {\bibfnamefont {K.}~\bibnamefont
  {Dusling}}, \bibinfo {author} {\bibfnamefont {G.~D.}\ \bibnamefont {Moore}},
  \ and\ \bibinfo {author} {\bibfnamefont {D.}~\bibnamefont {Teaney}},\ }\href
  {\doibase 10.1103/PhysRevC.81.034907} {\bibfield  {journal} {\bibinfo
  {journal} {Phys. Rev.}\ }\textbf {\bibinfo {volume} {C81}},\ \bibinfo {pages}
  {034907} (\bibinfo {year} {2010})},\ \Eprint {http://arxiv.org/abs/0909.0754}
  {arXiv:0909.0754 [nucl-th]} \BibitemShut {NoStop}%
\bibitem [{\citenamefont {Rapp}(2013)}]{Rapp:2013nxa}%
  \BibitemOpen
  \bibfield  {author} {\bibinfo {author} {\bibfnamefont {R.}~\bibnamefont
  {Rapp}},\ }\href {\doibase 10.1155/2013/148253} {\bibfield  {journal}
  {\bibinfo  {journal} {Adv. High Energy Phys.}\ }\textbf {\bibinfo {volume}
  {2013}},\ \bibinfo {pages} {148253} (\bibinfo {year} {2013})},\ \Eprint
  {http://arxiv.org/abs/1304.2309} {arXiv:1304.2309 [hep-ph]} \BibitemShut
  {NoStop}%
\bibitem [{\citenamefont {Song}\ \emph {et~al.}(2018)\citenamefont {Song},
  \citenamefont {Cassing}, \citenamefont {Moreau},\ and\ \citenamefont
  {Bratkovskaya}}]{Song:2018xca}%
  \BibitemOpen
  \bibfield  {author} {\bibinfo {author} {\bibfnamefont {T.}~\bibnamefont
  {Song}}, \bibinfo {author} {\bibfnamefont {W.}~\bibnamefont {Cassing}},
  \bibinfo {author} {\bibfnamefont {P.}~\bibnamefont {Moreau}}, \ and\ \bibinfo
  {author} {\bibfnamefont {E.}~\bibnamefont {Bratkovskaya}},\ }\href {\doibase
  10.1103/PhysRevC.97.064907} {\bibfield  {journal} {\bibinfo  {journal} {Phys.
  Rev.}\ }\textbf {\bibinfo {volume} {C97}},\ \bibinfo {pages} {064907}
  (\bibinfo {year} {2018})},\ \Eprint {http://arxiv.org/abs/1803.02698}
  {arXiv:1803.02698 [nucl-th]} \BibitemShut {NoStop}%
\bibitem [{\citenamefont {{Vorobyev, Ivan}}(2018)}]{Vorobyev:2018}%
  \BibitemOpen
  \bibfield  {author} {\bibinfo {author} {\bibnamefont {{Vorobyev, Ivan}}},\
  }\href {\doibase 10.1051/epjconf/201817118011} {\bibfield  {journal}
  {\bibinfo  {journal} {EPJ Web Conf.}\ }\textbf {\bibinfo {volume} {171}},\
  \bibinfo {pages} {18011} (\bibinfo {year} {2018})}\BibitemShut {NoStop}%
\bibitem [{\citenamefont {Contin}(2016)}]{Contin:2016vgc}%
  \BibitemOpen
  \bibfield  {author} {\bibinfo {author} {\bibfnamefont {G.}~\bibnamefont
  {Contin}} (\bibinfo {collaboration} {STAR}),\ }\bibfield  {booktitle} {\emph
  {\bibinfo {booktitle} {{Proceedings, 25th International Conference on
  Ultra-Relativistic Nucleus-Nucleus Collisions (Quark Matter 2015): Kobe,
  Japan, September 27-October 3, 2015}}},\ }\href {\doibase
  10.1016/j.nuclphysa.2016.02.064} {\bibfield  {journal} {\bibinfo  {journal}
  {Nucl. Phys.}\ }\textbf {\bibinfo {volume} {A956}},\ \bibinfo {pages} {858}
  (\bibinfo {year} {2016})}\BibitemShut {NoStop}%
\bibitem [{\citenamefont {Baym}\ \emph {et~al.}(2017)\citenamefont {Baym},
  \citenamefont {Hatsuda},\ and\ \citenamefont {Strickland}}]{Baym:2017qxy}%
  \BibitemOpen
  \bibfield  {author} {\bibinfo {author} {\bibfnamefont {G.}~\bibnamefont
  {Baym}}, \bibinfo {author} {\bibfnamefont {T.}~\bibnamefont {Hatsuda}}, \
  and\ \bibinfo {author} {\bibfnamefont {M.}~\bibnamefont {Strickland}},\
  }\href {\doibase 10.1103/PhysRevC.95.044907} {\bibfield  {journal} {\bibinfo
  {journal} {Phys. Rev.}\ }\textbf {\bibinfo {volume} {C95}},\ \bibinfo {pages}
  {044907} (\bibinfo {year} {2017})},\ \Eprint
  {http://arxiv.org/abs/1702.05906} {arXiv:1702.05906 [nucl-th]} \BibitemShut
  {NoStop}%
\end{thebibliography}%

\end{document}